\documentclass[manuscript,screen,authorversion,nonacm]{acmart}

\AtBeginDocument{%
  \providecommand\BibTeX{{%
    \normalfont B\kern-0.5em{\scshape i\kern-0.25em b}\kern-0.8em\TeX}}}

\setcopyright{acmcopyright}
\copyrightyear{2023}
\acmYear{2023}
\acmDOI{XXXXXXX.XXXXXXX}

\acmConference[Conference acronym 'XX]{Make sure to enter the correct
  conference title from your rights confirmation emai}{June 03--05,
  2018}{Woodstock, NY}
\acmPrice{15.00}
\acmISBN{978-1-4503-XXXX-X/18/06}

\usepackage{amsmath,amsfonts}
\usepackage{algorithmic}
\usepackage{graphicx}
\usepackage{textcomp}
\usepackage{xcolor}
\usepackage{natbib}
\usepackage{enumitem}
\usepackage{hyperref}
\usepackage{balance} 
\usepackage{multirow}

\makeatletter
\newcommand\footnoteref[1]{\protected@xdef\@thefnmark{\ref{#1}}\@footnotemark}
\makeatother

\makeatletter
\def\@copyrightspace{\relax}
\makeatother

\begin{document}

\title{A Systematic Review on Reproducibility in Child-Robot Interaction}


\author{Micol Spitale}
\authornote{Authors contributed equally to this research.}
\email{ms2871@cam.ac.uk}
\orcid{0000-0002-3418-1933}
\affiliation{%
  \institution{University of Cambridge}
  \country{UK}
}
\author{Rebecca Stower}
\authornotemark[1]
\email{stower@kth.se}
\affiliation{%
  \institution{KTH Royal Institute of Technology}
  \country{Sweden}
}
\author{Elmira Yadollahi}
\authornotemark[1]
\email{elmiray@kth.se}
\affiliation{%
  \institution{KTH Royal Institute of Technology}
  \country{Sweden}
}
\author{Maria Teresa Parreira}
\authornotemark[1]
\email{mb2554@cornell.edu}
\affiliation{%
  \institution{Cornell University}
  \country{USA}
}

\author{Nida Itrat Abbasi}
\email{nia22@cam.ac.uk}
\affiliation{%
  \institution{University of Cambridge}
  \country{UK}
}

\author{Iolanda Leite}
\email{iolanda@kth.se}
\affiliation{%
  \institution{KTH Royal Institute of Technology}
  \country{Sweden}
}

\author{Hatice Gunes}
\email{hg410@cam.ac.uk}
\affiliation{%
  \institution{University of Cambridge}
  \country{UK}
}

\renewcommand{\shortauthors}{Spitale, Stower, Yadollahi, Parreira et al.}

\begin{abstract}
Research reproducibility -- i.e., rerunning analyses on original data to replicate the results -- is paramount for guaranteeing scientific validity. However, reproducibility is often very challenging, especially in research fields where multi-disciplinary teams are involved, such as child-robot interaction (CRI). This paper presents a systematic review of the last three years (2020-2022) of research in CRI under the lens of reproducibility, by analysing the field for transparency in reporting. Across a total of 325 studies, we found deficiencies in reporting demographics (e.g. age of participants), study design and implementation (e.g. length of interactions), and open data (e.g. maintaining an active code repository). From this analysis, we distil a set of guidelines and provide a checklist to systematically report CRI studies to help and guide research to improve reproducibility in CRI and beyond. 
\end{abstract}

\ccsdesc[500]{Human-centered computing~HCI theory, concepts and models}
\ccsdesc[300]{Human-centered computing~Empirical studies in HCI}
\keywords{survey, child-robot interaction, reproducibility, replication, transparency}


\maketitle

\section{Introduction}
\label{sec:introduction}

The replication crisis -- also referred to as a ``crisis of confidence'' -- refers to a series of events circa 2015 that called into question the scientific validity of many psychological studies \cite{2020jost}, due to the difficulties in replicating apparently well-established psychological findings, as well as the differences in results between original and replicated studies.  Although initially emerging in psychology, many of the concerns raised, such as lack of open access to data, materials, and/or experimental design apply also to other (social) sciences. Among these are both Human Robot Interaction (HRI) and its related sub-field of Child Robot Interaction (CRI), where social and psychological relationships between humans and robots are often the focus of the research. Given its novelty and rapidly evolving progress, CRI in particular suffers from fragmented and heterogeneous literature, varying research goals, and a lack of standardised methods and metrics. 

Recent efforts have brought forth conversations related to replication specifically within CRI \cite{stower2023workshop,stower2021interdisciplinary}, with authors appealing for more works that address the main challenges in HRI with children whilst still ensuring high-quality reporting and data sharing. However, clear open science guidelines on reproducibility in HRI and related sub-fields are still missing. Consequently, the goal of this review is to systematically evaluate CRI works in the context of replicability. We aim to both provide an overview of the field as well as develop guidelines for CRI researchers.

In the following sections, we discuss both HRI and CRI in the context of the replication crisis. This is followed by the systematic review methodology (Section \ref{sec:methods}) and presentation and discussion of the results (Section \ref{sec:results}). From these, we developed a set of guidelines specifically tailored for CRI researchers. Finally, we discuss open challenges and future directions for the field of CRI as a whole (Section \ref{sec:discussion}).

\subsection{Replication Crisis}
The replication crisis was initially sparked in 2015 when a team of psychologists, led by Brian Nosek, published a paper in Science that provided evidence of a systemic issue with the reproducibility of psychological studies \cite{nosek2015promoting}. Specifically, they found that, across 100 studies, only 40\% were able to have their findings reproduced. The remaining experiments either did not replicate or produced mixed results. In addition, the replications that did succeed had less robust results than the original studies. In comparison to studies with weaker experimental findings, those that tended to replicate had more results that were highly significant originally. Since then, many efforts have been made to improve reproducibility in psychology, including preregistering studies, a priori power analyses (with the goal of encouraging larger sample sizes), and increased sharing of research materials \cite{nosek2022replicability}. However, those efforts have not yet been extended to other related fields such as human-robot interaction, where reproducibility is still very challenging.

\subsubsection{Replication in HRI}
Within the HRI literature, many researchers have already brought forth reflections and recommendations to address the issues of \textit{replicability} (reconducting the same experiment)  and \textit{reproducibility} (rerunning analyses on original data to replicate the results) \cite{lucking2018geographically, hoffman2020, gunes2022reproducibility, leichtmann2022crisis,cordero2022reporting}. These studies highlight that ensuring reproducibility is essential for building a strong knowledge base, identifying best practices, and fostering the development of effective and safe robotic systems. However, translating these recommendations into practice remains a strong barrier to much HRI research. 

One of the main factors that hinders reproducibility in HRI is the lack of transparent, rigorous reporting \cite{baxter2016characterising, veling2021qualitative, leichtmann2022crisis}. Recently, \citet{gunes2022reproducibility} published a review on reproducibility practices in HRI. Their review identifies technical challenges, research biases, and social systematic biases which can emerge in HRI research and pose threats to reproducibility. Technical challenges include, for example, the cost of commercially available robot platforms, and barriers to code/software sharing. Research biases discussed are null-hypothesis significance testing, sample size bias, positive results bias, and physical robot bias. Social systemic biases are classism bias, racial, ethnic, and cultural sampling (i.e., WEIRD populations), and robot morphology biases. They outline some of the practices that are being taken to address these issues, but stress that there is still much room for improvement within the field.

\citet{leichtmann2022crisis} also investigated the main contributors and potential solutions for the ``replicability crisis'' \cite{openscience} in HRI. They mention the effect of publication bias, which is rooted in the scientific system as the ``search for novelty''. The emphasis towards publishing only significant and/or novel research findings provides little incentive for researchers to engage in replicability/reproducibility studies, while also leading to questionable research practices (e.g., p-hacking) \cite{romero2019phylo}. Other causes include low-powered studies (due to small sample sizes and/or overly complex research designs) and the lack of a research framework that can be utilised across the field. Such lack of theory is reflected in disagreement even about what constitutes a successful replication \cite{scheel2021hypothesis}.

Nonetheless, recently a number of researchers have been providing positive examples of the importance of replicating HRI  studies.  \citet{ullman2021challenges} presented results of three studies targeted at investigating if being ``in the loop'', i.e., involved with a robot's actions, increases trust. They found that the initial, statistically significant effect failed to replicate once repeated with a larger sample size. 
Similarly, \citet{irfan2018social} also discussed the applicability of the replication crisis to HRI. They present findings where they failed to replicate a social facilitation effect with robots. They speculated about the reasons for their failures, such as the small effect sizes of social facilitation, the setting in which experimental data is collected, or a bias towards publishing only positive results. 
Other studies have made some effort in lowering barriers to reproducibility in HRI by proposing replication studies \cite{strait2020three}. For example, \citet{strait2020three} conducted a replication study wherein they performed the same experiment (investigating the joint-simon effect with a NAO robot) at three different locations. Whilst these works highlight some of the specific challenges that HRI faces, namely external validity – more studies are needed to understand just to what extent other variables, such as the experimental context, are impacting the results. 

\subsubsection{Replication in CRI}
This absence of research theory and well-established methods is particularly pronounced in the field of child-robot interaction \cite{belpaeme2013child}. CRI has become a rapidly growing research field in recent years, driven by the potential of robots to support children's learning \cite{belpaeme2013child}, development \cite{yadollahi2022motivating}, and well-being \cite{abbasi2022measuring, abbasi2022can}. Children are a diverse and dynamic population, with wide variations in cognitive and emotional development, as well as individual differences. This makes designing experiments that are both ecologically valid and controlled a difficult task \cite{stower2023workshop}. The relative novelty and rapid evolution of CRI as a research field contribute to its reproducibility challenges. Compared to HRI, CRI is still in its early stages, resulting in fragmented and heterogeneous literature with varying research goals and practices \cite{belpaeme2013child}. Consequently, whilst CRI inherits many of the challenges from HRI regarding reproducibility, it is not limited to them. 

First, \textit{recruiting the desired number and population for CRI studies} can prove to be especially challenging \cite{stower2023workshop}, and the inherent variability and diversity of children as a participant group presents challenges in designing controlled experiments that yield consistent results. In fact, children encompass a wide range of ages, developmental stages, cognitive abilities, and cultural backgrounds, making it difficult to generalise findings across different populations.

Second, children's engagement, preferences, and responses to robots can vary significantly, resulting in \textit{contextual dependencies} \cite{chen2020impact} that require careful consideration when attempting to replicate experiments. Children's behaviours and responses can be influenced by factors such as the appearance \cite{spitale2020whom}, behaviour \cite{zhang2023social}, and capabilities of the robot \cite{beran2011understanding}, as well as the specific task or activity being performed. Replicating these interactions across different contexts and settings often necessitates customisation and adaptation of the robotic platform and experimental protocols, further complicating reproducibility efforts.

Third, \textit{ethical considerations} play a critical role in CRI \cite{langer2023ethical}, posing additional challenges for reproducibility \cite{tolksdorf2021ethical}. Interactions between robots and children have the potential to impact children's attitudes, beliefs, and behaviours. As a result, researchers must navigate complex ethical issues surrounding privacy, autonomy, and the well-being of child participants. Adhering to ethical guidelines, obtaining informed consent from both children and their parents/guardians, and ensuring participant safety requires careful attention, potentially influencing the reproducibility of studies in CRI \cite{baxter2016characterising}. Such ethical considerations also affect the design stage of the studies, where researchers face trade-offs between research interests and ethical reflections (e.g., data collection vs. data privacy) but also constrains the sharing of experimental data \cite{baxter2016characterising}.

Lastly, CRI (as HRI), often involves \textit{multidisciplinary teams of researchers} from different domains, such as psychology, education, engineering, and design, which may have different assumptions, theories, and practices. This multitude of disciplines can introduce diverse perspectives and assumptions that need to be harmonised for replication efforts \cite{stower2021interdisciplinary}. Thus, ensuring reproducibility in CRI requires not only technical rigor but also interdisciplinary collaboration, transparency, and openness to alternative perspectives \cite{stower2023workshop}.

\subsection{Our Contribution}

In sum, CRI research faces empirical challenges related to the usability, deployment, and sustainability of long-term interactions, as well as ethical considerations of conducting user studies with children \cite{tolksdorf2021ethical, couto2022child}. This lack of standardised methods, metrics, and best practices can hinder the reproducibility of studies. As a result, CRI as a field is currently lacking specific guidelines and recommendations that may assist researchers in overcoming the above-mentioned challenges surrounding CRI and thus foster its reproducibility. To address this gap, we conducted a systematic review targeted at assessing current reproducibility standards within CRI, with the goal of proposing a set of guidelines for CRI researchers.

We conducted a systematic review based on PRISMA guidelines \cite{moher2009preferred} on the last three years of CRI research (2020-2022) to capture the most recent efforts of the CRI community with respect to reproducibility. We provide an overview of the field, identifying recent trends in robotic applications, robots used, the child populations involved, study design practices etc. This is followed by an analysis of the field in terms of transparency in reporting. We considered factors such as which study details were reported and to what extent they were described (e.g., whether the paper reported only the mean age of the sample but not the age range). 
Finally, from our findings, we distilled a set of guidelines  for enhancing reproducibility of CRI studies, while also discussing the main open challenges and future directions of reproducibility in CRI.

The main contributions of this work are the following:
\begin{itemize}
    \item We present the first systematic review that focuses on reproducibility and replication in child-robot interaction.
    \item We propose a set of guidelines to improve reproducibility in CRI (which can be further extended to HRI or HCI) by discussing the main challenges that emerged;
    \item We provide a checklist to systematically report child-robot interaction studies to help and guide researchers in improving reproducibility of their works;
    \item We draw a research agenda by highlighting the main open challenges in the field of CRI in terms of reproducibility.
\end{itemize}

\section{Methods}
\label{sec:methods}

\begin{figure}[htb!]
    \centering
    \includegraphics[width=\textwidth]{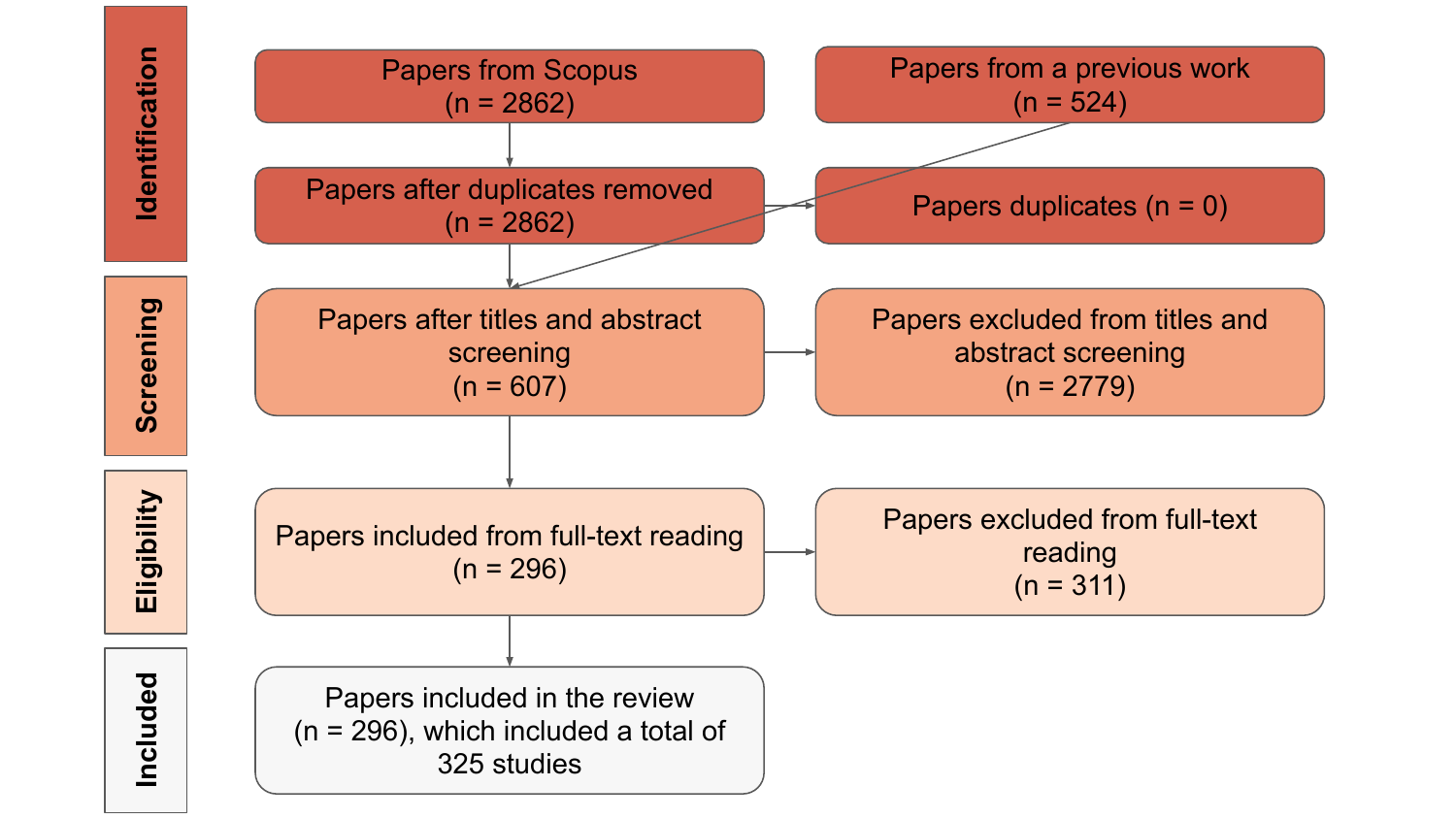}
    \caption{PRISMA schema steps.}
    \label{fig:prisma}
\end{figure}

This section describes the methodology adopted to conduct our literature review, including the screening procedure (Section \ref{sec:screen-proc}) and the description of data extraction and analysis (Section \ref{sec:data-extr}). 
We pre-registered this work\footnote{\url{https://osf.io/h8bka/?view_only=04088596f64846229b236cb962d9ea8c}} via the OSF platform following all the guidelines of Open Science \cite{openscience}.

\subsection{Screening Procedure} 
\label{sec:screen-proc}
We followed the guidelines described in \citet{nightingale2009guide} to conduct this systematic review. We used the PRISMA framework \cite{moher2009preferred} as it represents the state-of-the-art workflow for systematic reviews and it includes the following steps (as depicted in Figure \ref{fig:prisma}):
\begin{enumerate}
    \item \textbf{identification}: searching specific databases (how) within a specific time range (when);
    \item \textbf{screening}: filtering out manuscripts whose title and abstract do not meet the eligibility criteria; 
    \item \textbf{selection}: reading the full text of the manuscripts and excluding the ones that do not meet eligibility criteria;
    \item \textbf{inclusion}: selecting the final list of the included manuscripts whose data are extracted.
\end{enumerate}

\subsubsection{Search Query}
For generating the list of manuscripts to include in this survey, we re-used a list of already screened manuscripts that were extracted via Scopus from January 2020 to December 2021 in another work. Then, we identified the newest manuscripts from January 2022 to July 2022 using the same approach (i.e., the same database and search query).
We searched for these manuscripts via the Scopus database (as in \citet{catania2022conversational}). The list of terms for the search included both robot technology and the children population. 
The manuscripts were exported on 12/07/2022 and the search query used was the following:
\begin{verbatim}
TITLE-ABS-KEY  ("school"  OR  "education"  OR  "rehabilitation")
AND  ("child*"  OR  "teen*") AND  "robot" 
AND  ("study"  OR  "experiment*"  OR  "field study"  
OR  "into the wild") AND  PUBYEAR  =  2022.
\end{verbatim}

After the extraction, we filtered out the duplicates and stored references in an Excel file, publicly available in the GitHub repository\footnote{\label{github-repo}\url{https://github.com/becbot/CRI_Replication}}.

\subsubsection{Eligibility Criteria} 

We defined a set of inclusion and exclusion criteria as follows. 
The papers were included if:
\begin{itemize}
    \item they included a social interaction between a child and a robot;
    \item they reported either physical or online studies, as long as they met all other criteria;
    \item they described some evaluation by the child (i.e., not only teachers/ parents/ healthcare providers, etc).
\end{itemize}

Papers were excluded if:
\begin{itemize}
    \item they were review/theoretical papers;
    \item they included only adults (older than 18 years old);
    \item they were purely technical (e.g., medical robotics).
\end{itemize}

\subsubsection{Selection Process} 
\makeatletter

\makeatother
All the manuscripts extracted from the database were first screened by reading their titles and abstracts. Then we conducted a more in-depth analysis of the full texts to ensure that the manuscripts met the eligibility criteria. 
Before starting the screening process, a random sample of 20 papers was taken from the 2022 Scopus search and screened by 5 reviewers to ensure consensus on inclusion/exclusion criteria. All other papers were randomly divided by the reviewers and screened by one person. Titles and abstracts were screened based on the eligibility criteria mentioned above. If a reviewer was unsure about a paper, they flagged it for further discussion between all reviewers. The same reviewers were then assigned to a different subset of the manuscripts for full-text screening and data extraction. 

We collected a total of 2862 papers using the search query (See Figure \ref{fig:prisma}). Then, we screened the manuscripts based on their titles and abstracts and selected a total of 607. After that, we added 524 manuscripts from a previous screening and reviewed the full texts. This resulted in a final number of 296 manuscripts (reported in the Supplementary Materials). Among the 296 manuscripts, we identified a total of 325 studies (i.e., some of the papers reported more than one study). 
When one of the reviewers was uncertain about the inclusion of a paper, it was assigned to one of the other reviewers to make a final decision. The final list of the extracted papers can be found in a publicly available spreadsheet stored in the GitHub repository\footnote{\label{github-repo}\url{https://github.com/becbot/CRI_Replication}}.

\subsection{Data Extraction and Analysis}
\label{sec:data-extr}

We identified a set of variables that can inform the transparency/ reproducibility of the included papers and extracted the corresponding data. Note that some variables were extracted on a paper level (e.g., author discipline, application context), whereas others were extracted on a study level (e.g., number of participants, robot model). See Table \ref{tab:variables} for a summary of the extracted/coded variables. We only extracted data from the content of the main paper and did not cover the supplementary material (if present), so as to follow the same extraction procedure across all papers (including those without any extra materials). However, we did note if supplementary material, appendices, or links to external repositories were included in the paper. 

\begin{table}[]
\footnotesize
    \centering
    \resizebox{\textwidth}{!}
   {
    \begin{tabular}{llllll}
            \toprule
\textbf{Variable}                          & \textbf{Category}                                & \textbf{Type}        & \textbf{Example}       & \textbf{Level}                                     &  \\
      \midrule
\multirow{5}{*}{General}          & Authors' discipline                      & categorical & Computer                                      science, Psychology, etc &  Paper\\
                                  & Country                                & categorical & USA, Finland, Japan, etc.                      & Study \\

                                  & Open access                             & categorical & Yes, No                                       & Paper \\
                                  & Ethics approval                         & categorical & Yes, No                                        &  Paper\\
                                  & Consent                                 & categorical & Yes, No                                        &  Paper\\
                                  & Preregistration                         & categorical & Yes, No                                        &  Paper\\
                                  \midrule
\multirow{7}{*}{Open Science}     & Open materials                          & categorical & Yes, No, Partially available                   & Paper \\
                                  & Open measures                           & categorical & Yes, No, Partially available                   &  Paper\\
                                  & Open code                               & categorical & Yes, No, Partially available, N/A   &  Paper\\
                                  & Code repository                         & categorical & Yes, No, Partially available, N/A   &  Paper\\
                                  & Open analysis                           & categorical & Yes, No, Partially available, N/A   & Paper \\
                                  & Open data                               & categorical & Yes, No, Partially available, N/A   &  Paper\\
                                  & Study repository                        & categorical & Yes, No, Partially available, N/A   &  Paper\\
                                  \midrule
\multirow{12}{*}{Design}          & Application scenario                    & categorical & Education, Therapy, Play, ...                  & Study \\
                                  & Number of sessions                      & numerical   & 1, 2, 10, ...                                      & Study \\
                                  & Length of single interaction & numerical   & 5 minutes, 10 minutes, 23 minutes ...                     &  Study\\
                                  &  with robot &    &    &  \\
                                  & Population                              & categorical & Neurotypical, ASD, Mixed, ...                 &  Study\\
                                  & Age range                               & numerical   & 4-5 years-old                                  &  Study\\
                                  & Mean age                                & numerical   &  10.26, 6.3, 7.7,...  &  Study\\
                                  & Gender reporting                         & categorical & Yes, No                                        & Study \\
                                  & Age and gender reporting & categorical & Yes, No                                        & Study \\
                                  & after exclusion &  &  &  \\

                                  & Number of participants                  & numerical   & 10, 25, ...                                    &  Study\\
                                  & Results reporting                        & qualitative & Figure, in-text, Not reported                  &  Study\\
                                  & Analyses reported                       & qualitative & Quantitative, Qualitative, Not reported        &  Study\\
                                  \midrule
\multirow{3}{*}{Robot/Technology} & Robot model                             & categorical & NAO, Cozmo, Misty, ...                         &  Study\\
                                  & Mode of interaction      & categorical & Real-life, Video, Life-mediated                &  Study\\
                                  & Robot operation                         & categorical & Autonomous, Wizard of Oz,     & Study\\
                                  &                          &  & Semi-autonomous      & \\
        \bottomrule
    \end{tabular}}
    \caption{List of variables extracted for each paper, the corresponding category, type, and example of values. N/A refers to "Not Applicable".}
    \label{tab:variables}
\end{table}

We first extracted general information about the paper (citation, the disciplines of the authors, and the country(s) in which the study was conducted). Author discipline was coded based on the departmental affiliations reported in the paper. However, we acknowledge that this method is imperfect, as researchers' departments do not necessarily have to correspond with their research backgrounds. Additionally, whilst some papers report departments, divisions, or even specific labs to which the authors belong, others report only the university affiliation/s, making it sometimes difficult to identify a specific discipline. 

We identified whether the paper reported obtaining ethical approval (0,1) and informed consent (0,1). We extracted these two variables separately, as it is possible in some cases that ethical approval is not required. Whether the paper was open-access (0,1) and had pre-registered hypotheses (0,1) was also recorded. 

Next, we extracted data related to the experimental design. First, two of the authors coded the application context (i.e., if there was a specific application such as healthcare or education that the study fit within). The different categories of application scenarios were decided post-hoc based on discussion between the two coders. Where studies could potentially fit multiple applications (e.g., \cite{liu2022auxiliary} could be both physical healthcare and education), the most relevant application was chosen, such that each paper was only coded to one application. The population/s being tested, and the number of participants were also extracted. If no specific population was reported (e.g., children with ASD), we coded the population as ``neurotypical''. We also extracted demographic information of the participants, such as the mean age, minimum and maximum age, and gender breakdown of participants. We further noted whether age and gender information was reported prior to or after any exclusion of participants.  

As studies in child-robot interaction can vary both in the number of interactions the child has with the robot and the length of these interactions, both of these were extracted from the papers. Number of interactions was coded as ``single", ``multiple", or ``not reported". However, we found it challenging to code the exact length of the interaction, especially when multiple interactions with the robot were considered. It was not always clear whether the lengths reported referred to the entire study session (i.e., including questionnaires, introduction, etc.) or only the interaction with the robot. Additionally, some authors reported the range of the interaction's duration (e.g., 15-20 minutes), whilst others reported the average time, making it even more difficult to consistently code the manuscripts.  Hence, we only coded the length of the interaction as reported (1) or not reported (0).

The type of interaction is also relevant in CRI, as some studies have shown that video and real-life interactions are not always comparable \cite{bainbridge2011benefits}. Consequently, we recorded whether the interaction occurred in ``real life", was ``live-mediated" (e.g., via video call), or used ``videos" or ``pictures" of robots as stimuli. Studies in which it was not possible to tell the mode of interaction were coded as ``not reported''. 

Similarly, as there is no single definition of what constitutes a ``social robot'', we considered also the robot model being used in the study, as well as how the robot was controlled during the interaction. The robot model was extracted directly from the papers. Papers that used their own custom robot were coded as ``custom'', otherwise if no robot model was mentioned this was classed as ``not reported''. For robot operation, we identified five different modes of operation: ``autonomous'' (no experimenter needed to operate the robot during the interaction), ``semi-autonomous'' (some experimenter input is needed at various stages), ``pre-scripted" (robot operated independently, but following a defined set of interaction steps), ``Wizard of Oz'', where a human operator completely controlled the robot, and ``teleoperated'', which is where the participant themselves controls some elements of the robot's behaviour. We classified the mode of operation based on the authors' own definitions/descriptions of the interaction, and if none was provided we coded this as ``not reported''. This means there are potential inconsistencies in coding between the papers, as different authors may define robot autonomy in different ways.  

Finally, we coded a set of variables relating to the open-science practices of the papers. First, we checked whether the paper had open materials, i.e., if it would be possible to replicate the task or interaction, given access to all of the materials in the information provided\footnote{Note that this does not consider how the materials are actually  obtained (e.g., cost, commercial availability), we only took into account if the materials are reported in sufficient detail that someone with access to them could replicate the study.}. Second, we coded whether the study had open measures, if applicable (that is, if it provided the materials used such as questionnaires, interview protocols, coding schemes, etc.). We distinguished between materials (stimuli or equipment needed to set-up and run the experiment, such as specific tasks or games) and measures (tools needed to capture and record data). Next, we coded whether the study had open code (if they provided the code used to run the robot, for instance) and/or a code repository. We separated these two variables as, in theory, a study could include all the code necessary to replicate the study within the paper, or could use entirely pre-existing code bases without needing their own repository. If the study used picture or video stimuli and thus could be replicated without needing any code, this was coded as not applicable. We then coded whether the data were open access, as well as if there was any data repository associated with the paper. Papers that explicitly stated that data could not be made available were coded as ``not applicable". We classified study repositories as ``active" (e.g., an external link to GitHub or OSF), ``inactive" (broken or dead links), ``supplementary", ``in-text" (all data was reported in text), or ``available upon request" if an explicit data availability statement was made. Table \ref{tab:variables} collects the list of variables extracted for each paper and the corresponding data type (i.e., numerical, categorical, or qualitative), as well as if data were extracted on the paper or study level.

\section{Results and Discussion}
\label{sec:results}

Following data extraction, we performed descriptive analyses on the included manuscripts (frequency, mean and SD) grouped by variable classes (i.e., general, open science, design, and robot/technology).

\subsection{General Overview} 
We first analysed the geographical distribution of the included papers, the disciplines of the authors, as well as the application context of the papers. 

\subsubsection{Geographical Distribution}

\begin{figure}
    \centering
    \includegraphics[width=\textwidth]{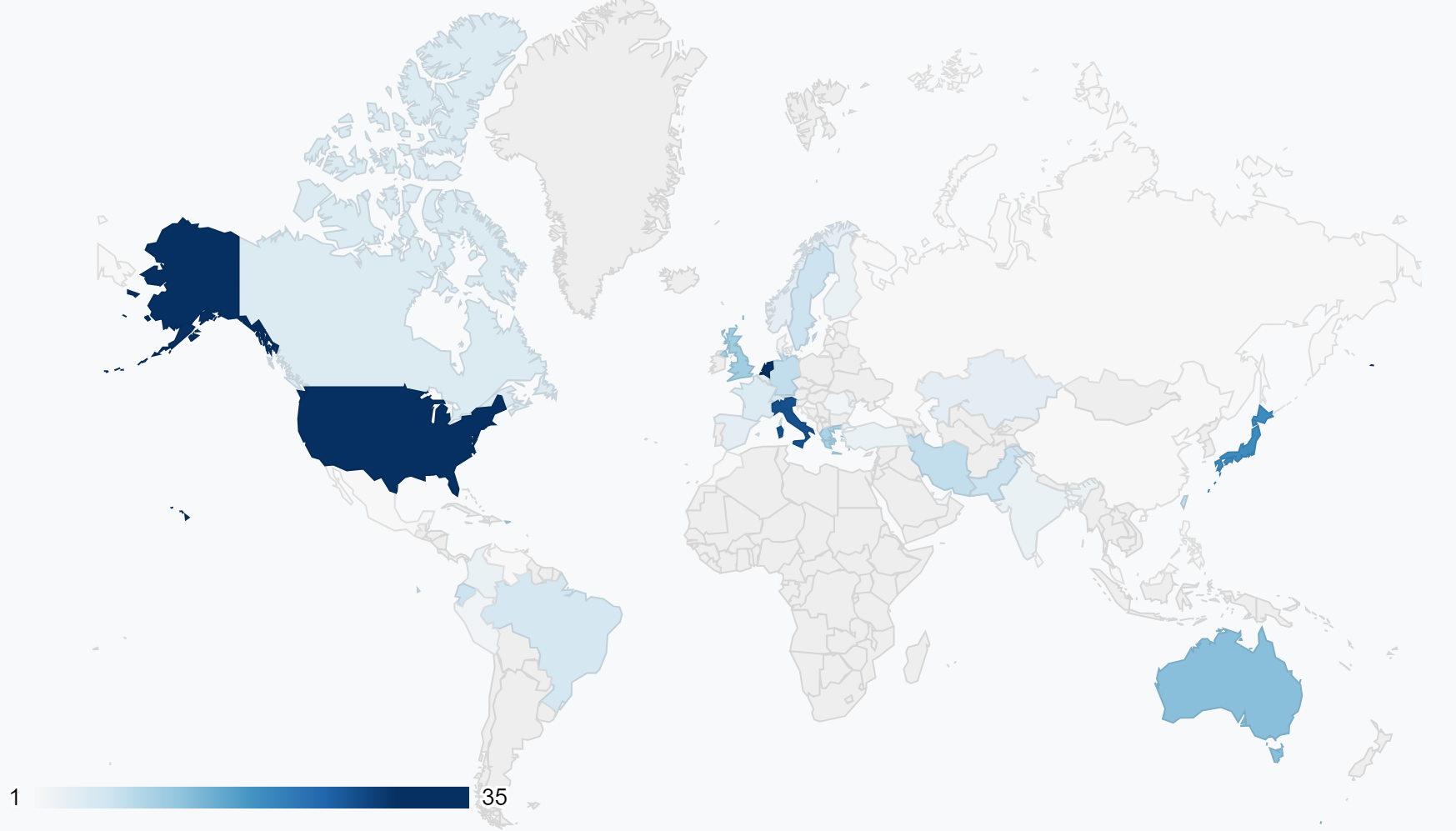}
    \caption{Map of the distribution of the studies across countries worldwide.}
    \label{fig:world-map}
\end{figure}

Within the surveyed manuscripts, we extracted the countries where each study was conducted, as depicted in Figure \ref{fig:world-map}. Out of the 256 included manuscripts, 6.08\% did not report where their study/s were conducted. The country with the highest number of reported studies was the US, accounting for 11.49\% of all manuscripts. 3 papers reported data collection across multiple countries. 
From a reproducibility standpoint, reporting where the study was conducted complemented reporting relevant ethical procedures. Reporting which ethical guidelines were followed also benefits researchers in decomposing what type of ethical standards were in place and how these could transfer to a different country's ethical guidelines. 

The majority of studies were conducted within the Western world, confirming other HRI works which have highlighted the disproportionate number of studies conducted with WEIRD populations \cite{gunes2022reproducibility}. Replication efforts across different countries, cultures, and demographics are needed to help understand to what extent findings relevant to CRI are generalisable across children's development, versus tied to specific contexts. 

\subsubsection{Author Disciplines}
\begin{figure}
    \centering
    \includegraphics[width=\textwidth]{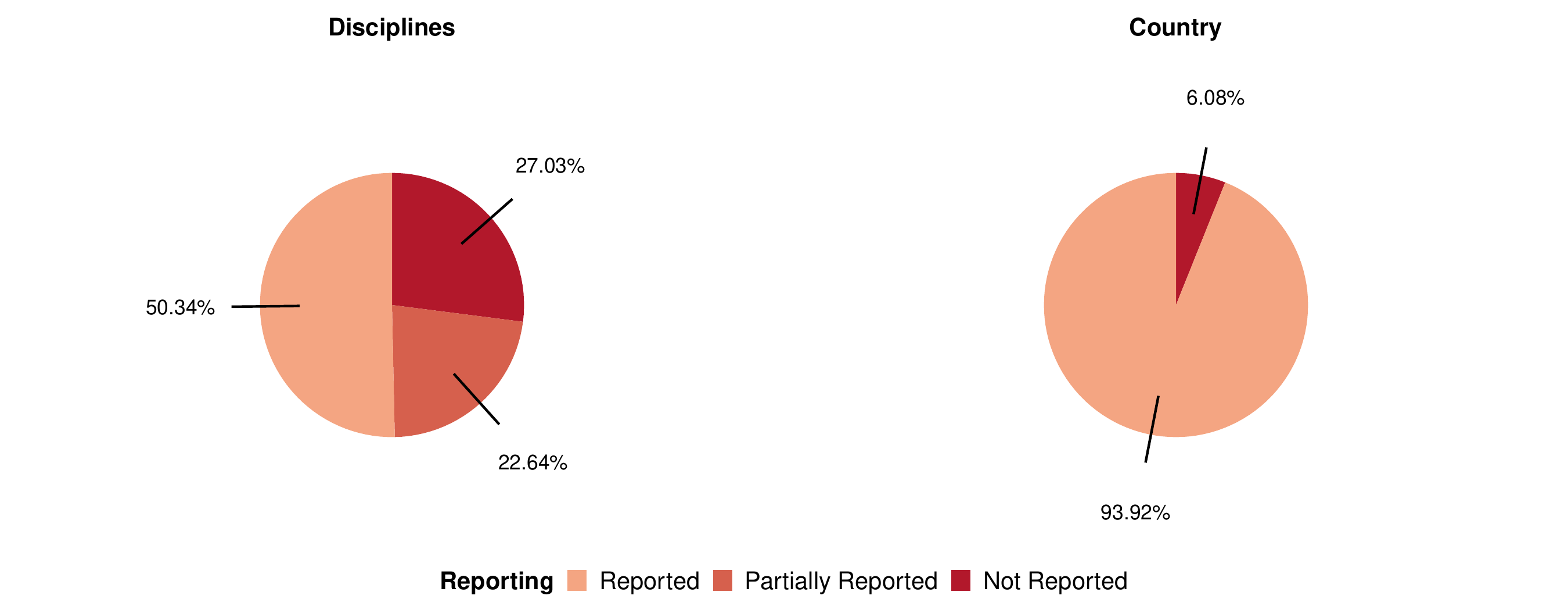}
    \caption{Reporting of country and discipline information across the included papers.}
    \label{fig:overview}
\end{figure}

\begin{figure}
    \centering
    \includegraphics[width=0.8\textwidth]{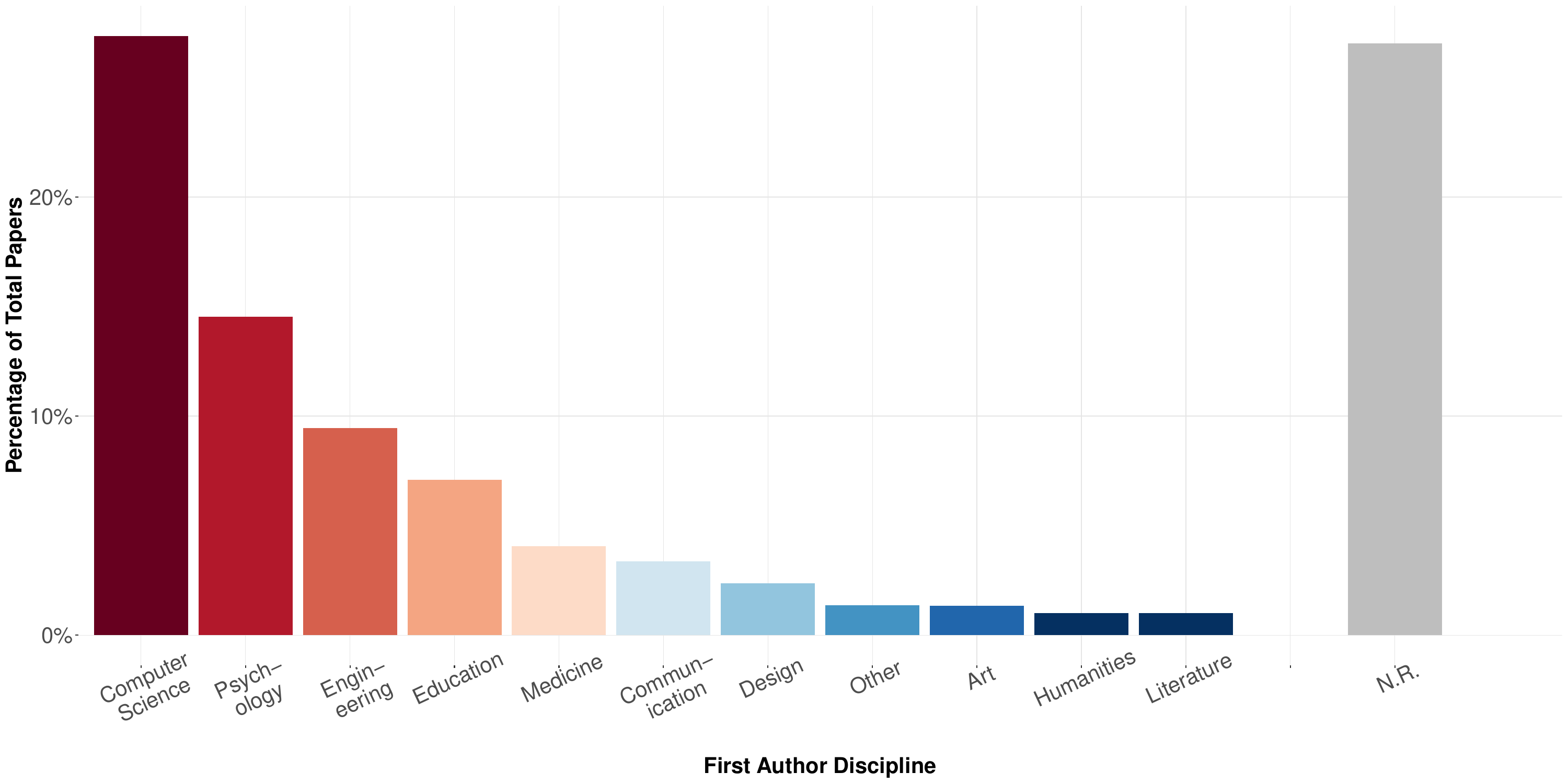}
    \caption{Frequency distribution of disciplines across the first authors in the included papers. Disciplines which occurred less than 1\% of the time across the dataset were coded as ``Other''. }
    \label{fig:first_disciplines}
\end{figure}

\begin{figure}
    \centering
    \includegraphics[width=0.8\textwidth]{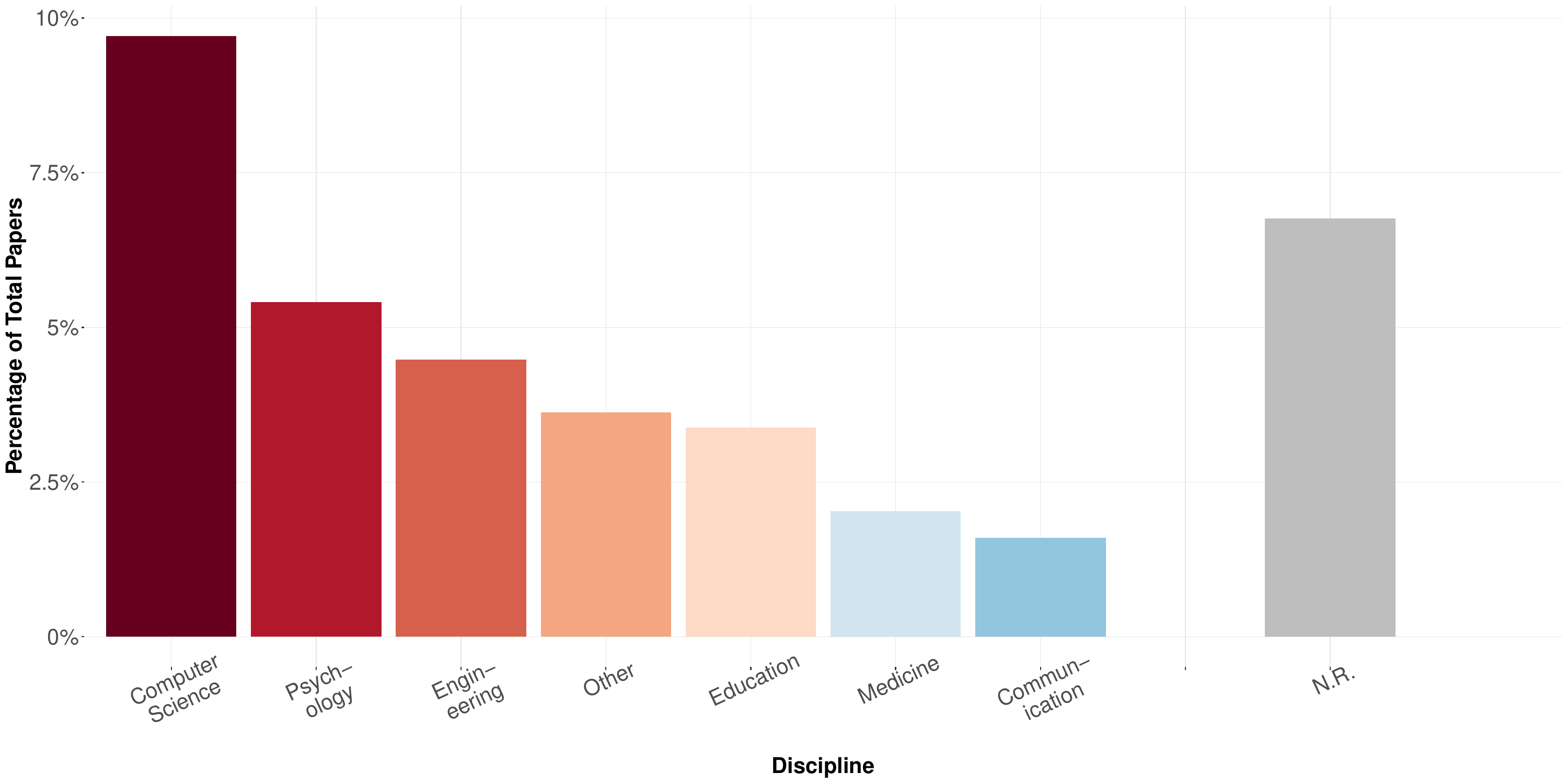}
    \caption{Frequency distribution of disciplines across all authors in the included papers. Disciplines which occurred less than 1\% of the time across the dataset were coded as ``Other''.}
    \label{fig:all_disciplines}
\end{figure}

We checked whether the authors’ disciplines were reported in the included manuscripts (Figure \ref{fig:overview}). 50.34\% of papers fully reported the authors’ disciplines, 22.64\% partially reported them (e.g., the disciplines reported covered just part of the authors), and 27.03\% of papers did not report the disciplines at all. 
The diversity of the disciplines (Figures \ref{fig:first_disciplines} and \ref{fig:all_disciplines}) attests to the interdisciplinary nature of the field. 
Different disciplines value different styles of reporting, thus enhancing the inconsistency we observe in reporting of CRI research. 
Added to this, a lack of reporting homogeneity among different fields and publishing venues imposed some limitations in extracting the authors' disciplines, confirming the need for better reporting guidelines targeted at interdisciplinary fields such as HRI, HCI, and CRI.

\subsubsection{Applications}
\begin{figure}
    \centering
    \includegraphics[width=0.8\textwidth]{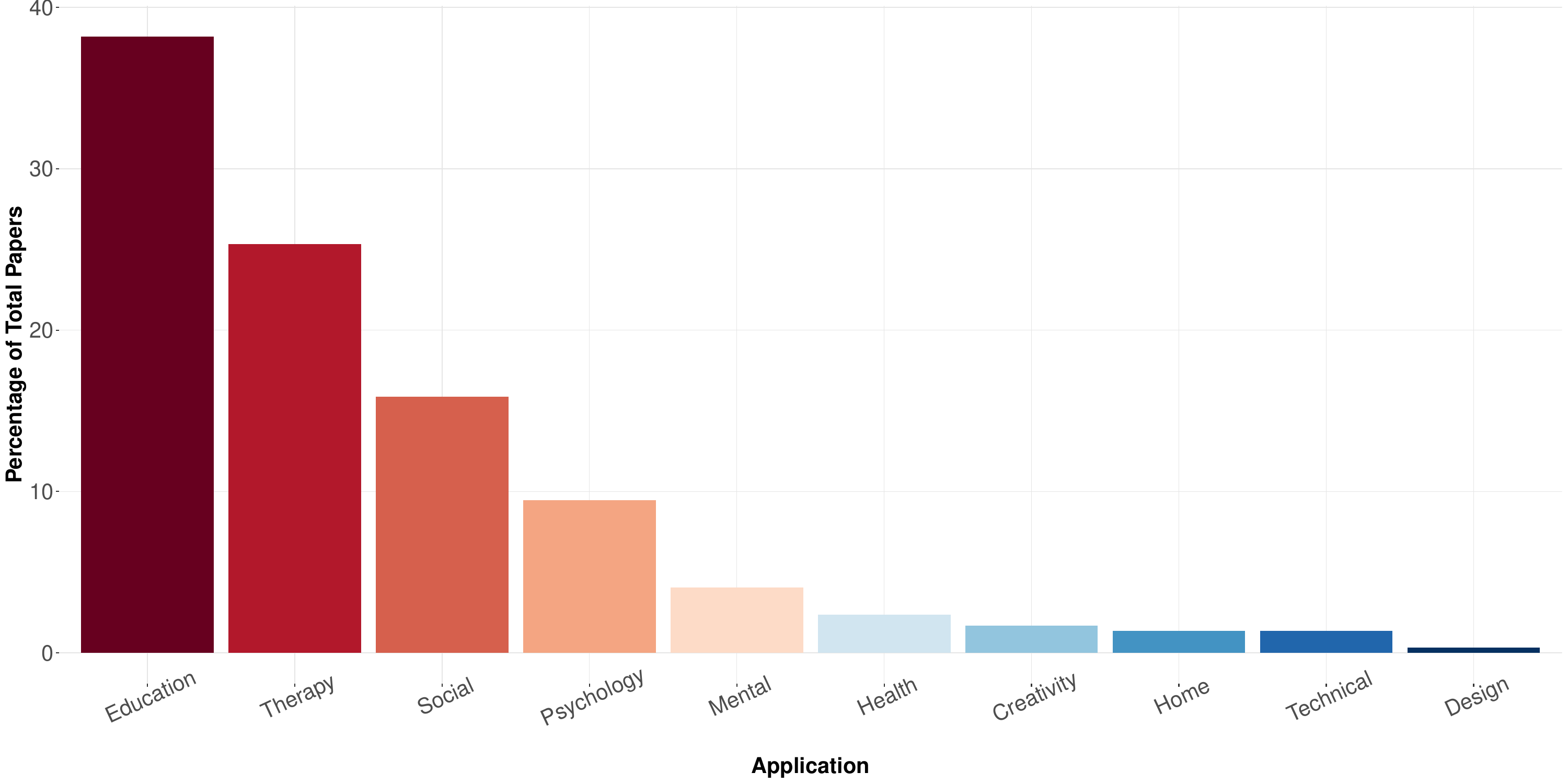}
    \caption{Frequency distribution of application contexts of the included papers. Each paper was coded to one application.}
    \label{fig:applications}
\end{figure}

From the surveyed papers 10 different application scenarios were identified (Figure \ref{fig:applications}). 38.18\% of the surveyed manuscripts reported an educational application, 25.34\% reported a therapeutic application, 15.88\% reported a social application (e.g., \cite{martin2020young} investigated children's helping behaviour towards a humanoid robot), 9.46\% reported a developmental psychology application (e.g., \cite{wang2020infants} examined infants' perceptions of cooperation involving robotic agents), 4.05\% reported a mental health application, 2.36\% reported a health application, 1.69\% reported a creativity application, 1.35\% reported a home application, 1.35\% reported a technical application, and 0.34\% reported a design application.

These results highlight a tendency towards high-stakes scenarios, namely health-related applications (as well as education-based studies). This is yet another reason why establishing guidelines for reproducibility in CRI is pertinent, as it requires additional ethical and practical considerations that need to be planned a priori, but also reported on adequately \cite{tolksdorf2021ethical}. Nevertheless, extracting the applications of the surveyed paper is not always straightforward, while some posit their research in a dedicated application, others aim for more application-agnostic reporting.

\subsection{Demographics}
Next, we considered the sample demographics across the included papers, including the targeted population/s, age and gender distributions, and sample size. 

\subsubsection{Population}
\begin{figure}
    \centering
    \includegraphics[width=0.8\textwidth]{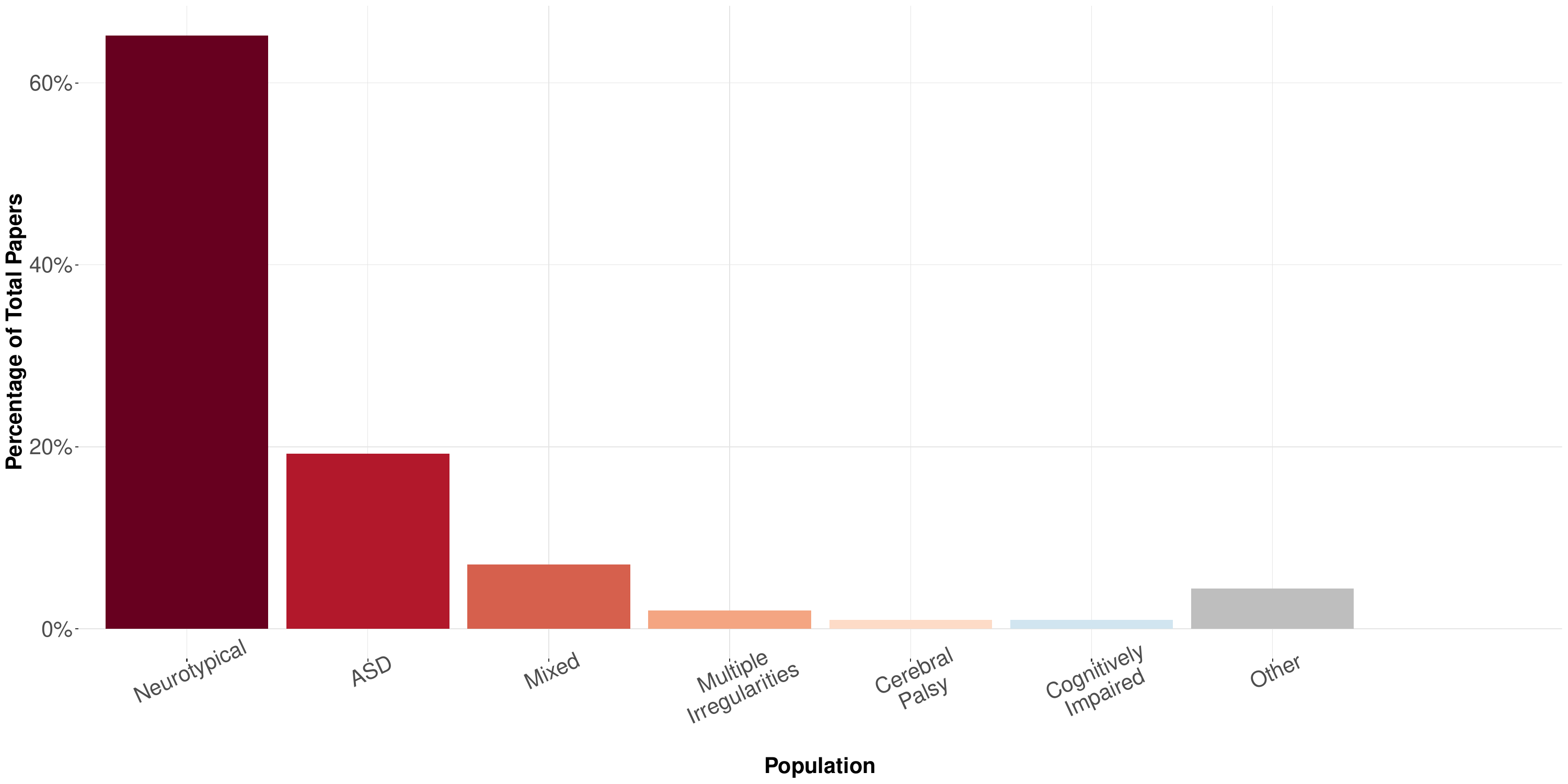}
    \caption{Frequency distribution of child populations targeted in the included papers. Populations which occurred less than 1\% of the time were coded as "Other"}
    \label{fig:gg_population}
\end{figure}

The majority of the surveyed papers, 64.86\%, included neurotypical children. 19.26\% reported that they included children with autism spectrum disorder (ASD) and 7.09\% reported that they included a mixture of populations (e.g., \cite{clabaugh2020month} involved both neurotypical children and children with autism in a one-month in-home human-robot interaction study). 2.36\% of the surveyed papers reported that they included a population with multiple diagnoses. 1.01\% of studies reported that their population included children with cerebral palsy, 1.01\% included children with cognitive impairments, 0.68\% included deaf children, and 0.68\% reported including hospitalised children. The remainder of the studies included one each of the following populations: children with behavioural disorders, cystic fibrosis, diabetes, down syndrome, dysgraphia, language disorders, and physical disability (Figure \ref{fig:gg_population}). 

A high percentage (35.14\%) of the surveyed works focused on non-neurotypical populations or other conditions. Recruiting and working with these populations is challenging \cite{stower2023workshop, yadollahi2021children}, leading researchers to have smaller participant pools, exclude many participants, and have shorter-duration studies. All of these factors play into reproducibility \cite{leichtmann2022crisis}. With vulnerable populations and high-stakes application scenarios, the need for literature that appropriately reports on the recruitment process and challenges, as well as provides a clear description of the study design and procedure, becomes obvious.

\subsubsection{Sample characteristics and reporting}
\begin{figure}
    \centering
    \includegraphics[width=\textwidth]{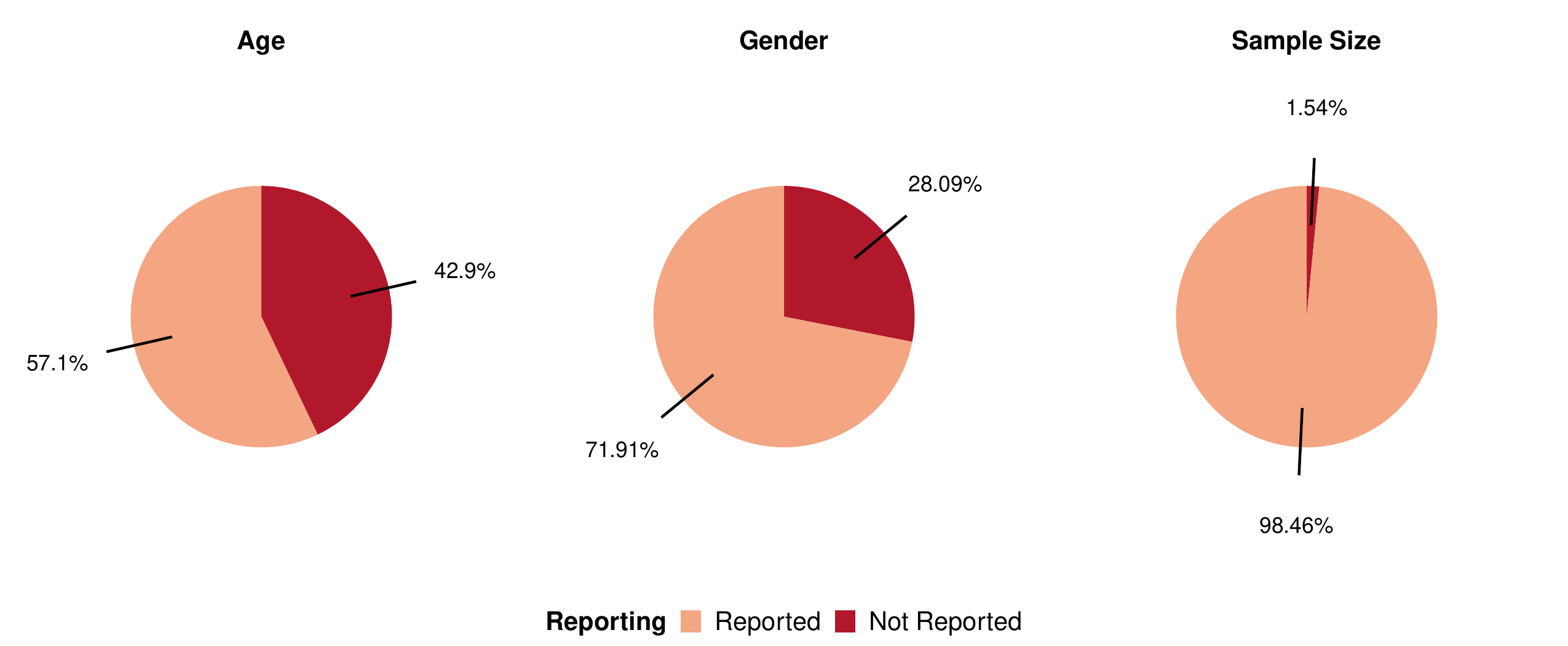}
    \caption{Percentage of studies reporting age, gender, and sample sizes.}
    \label{fig:demographics}
\end{figure}

We checked whether the authors reported the data about the age and gender of the participants before and after exclusion in their studies. 45.85\% of the included studies did not report any screening procedure to include or exclude the participants in their studies. Of the remaining studies, 21.54\% reported a screening procedure but did not report if the age and gender data included in the paper referred to the participants after the screening procedure or not. 24.31\% reported the age and gender information only after the screening, whereas 8.31\% reported age and gender only before the screening process. Zero of the surveyed papers reported age and gender both before and after screening. More details about reporting across the surveyed papers can be seen in Figure \ref{fig:demographics}.

Our results suggest that authors often reported participants' age and gender without referring to recruitment strategies and exclusion criteria used. 
To promote reproducibility and transparency practices, authors should report details of the sample after exclusion or mention if participants were excluded, and why. This is particularly critical in specifying if the reported age and gender correspond to the participant group before or after the exclusion.  

\begin{figure}
    \centering
    \includegraphics[width=0.6\textwidth]{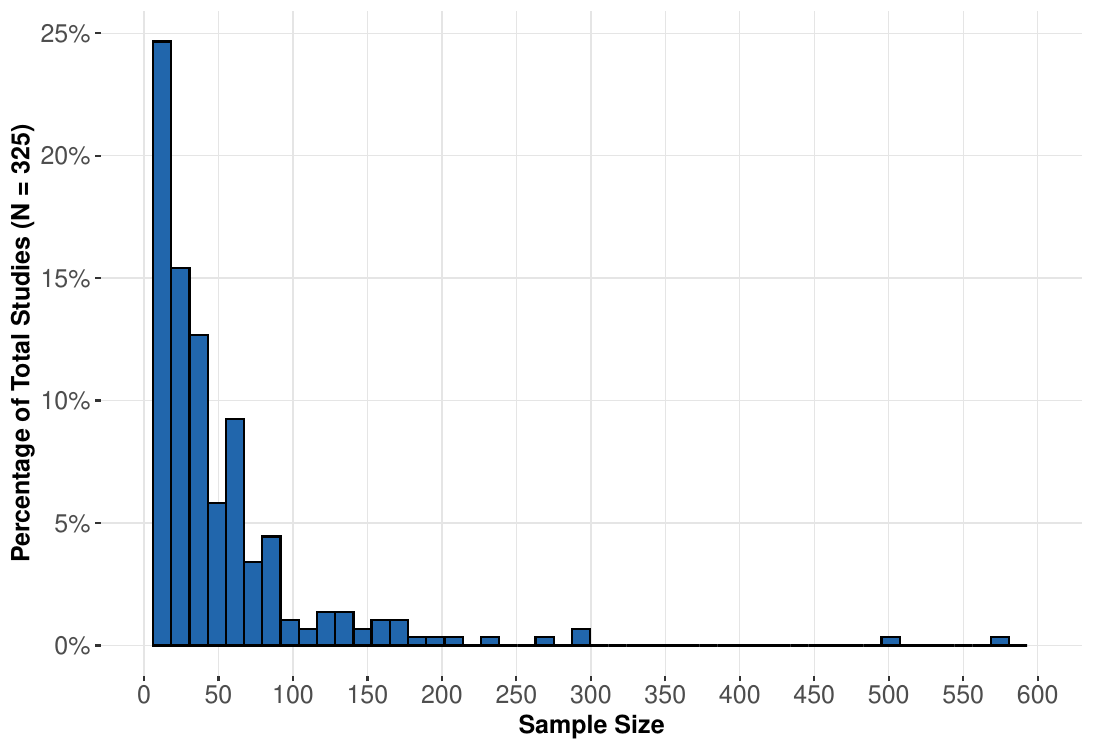}
    \caption{Distribution of sample sizes across the included studies.}
    \label{fig:sample_sizes}
\end{figure}

\begin{figure}
    \centering
    \includegraphics[width=0.6\textwidth]{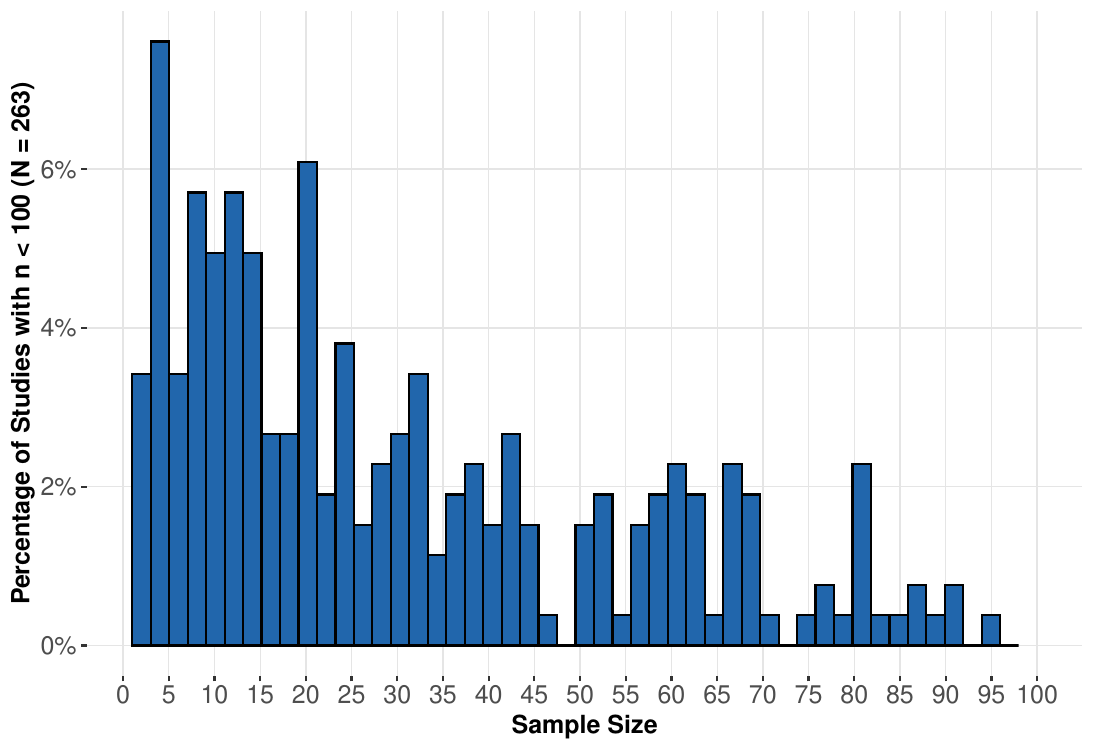}
    \caption{Distribution of sample sizes across studies with $n <100$.}
    \label{fig:sample_sizes2}
\end{figure}

The median sample size across all studies was $N = 25$ $(min = 1, max = 570)$. 6 studies (1.85\%) did not report the sample size. Although appropriate sample size depends on the specific type of experiment (e.g., case study, randomised control trial) and design (number of variables, expected effect size), the overall low median sample size suggests a trend towards likely underpowered studies within CRI. This is problematic from a reproducibility perspective, as underpowered studies both reduce the likelihood of a truly significant effect being detected (Type II error / false negative) but also increase the chance of a Type I error (false positive) occurring when significant effects \textit{are} present \cite{button2013power}. Figures \ref{fig:sample_sizes} and \ref{fig:sample_sizes2} detail the sample size distribution.


\subsubsection{Age}

\begin{figure}
    \centering
    \includegraphics[width=0.6\textwidth]{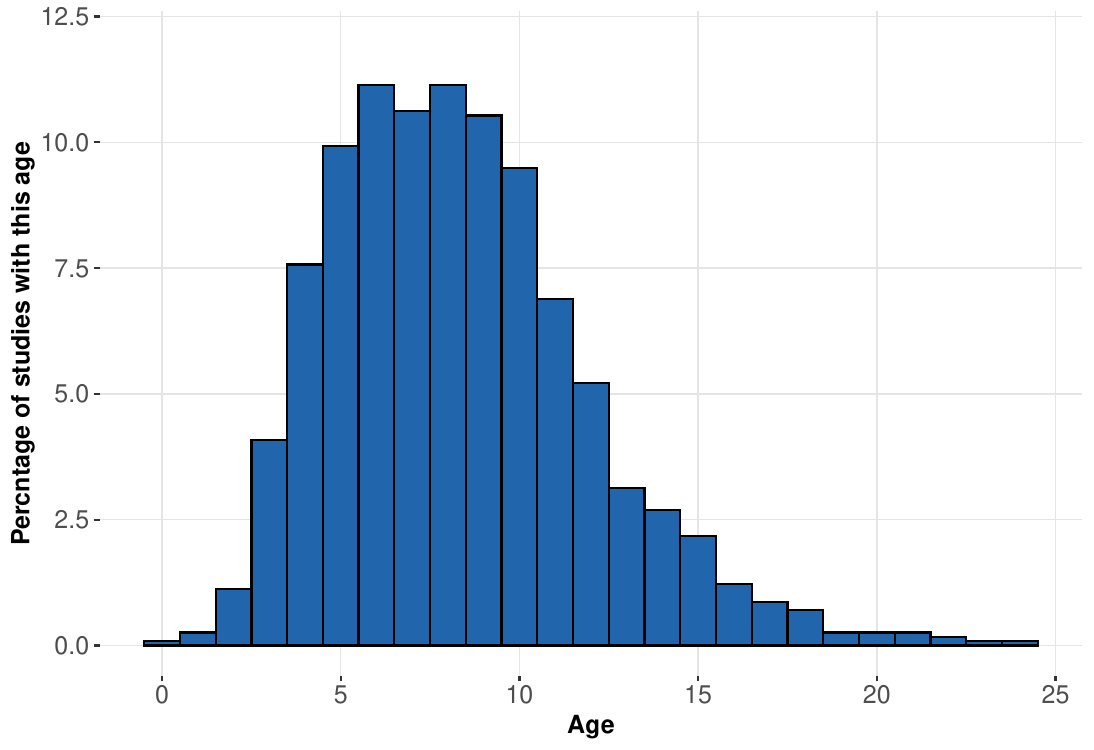}
    \caption{Frequency distribution of ages of children in the included papers.}
    \label{fig:gg_age1}
\end{figure}

The majority of the studies included in the manuscripts (75.69\%) reported either the age range (minimum and maximum age) or mean age, while 24.31\% of studies did not report either. The median age across all studies which reported age information was 7.65 y.o., with a mean age of 7.69 y.o., a maximum age of 17.4 y.o., a minimum age of 0.77 y.o., and a variance of 9.47 years. Figure \ref{fig:gg_age1} depicts the frequency distribution of age across all the studies included in this survey.

\subsubsection{Gender}
\begin{figure}
    \centering
    \includegraphics[width=0.6\textwidth]{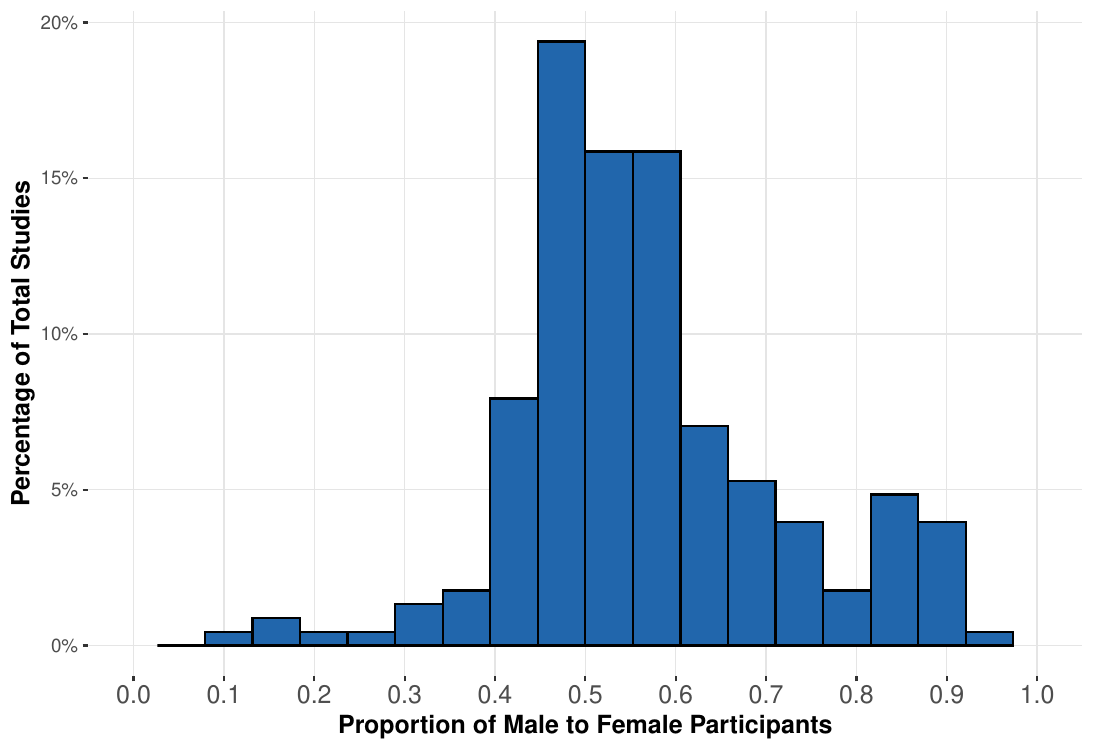}
    \caption{Distribution of the male-to-female participant ratio across the included papers. Higher proportions refer to a higher number of males.}
    \label{fig:gg_gender}
\end{figure}

Regarding gender, 71.69\% of the included studies reported the gender of the children involved in their studies, while 28.31\% did not report their participants' gender. 
Out of the reported ones, 47\% of participants were female and 53\% were male. Only one study reported including non-binary participants.  The average ratio of female to male participants, across all studies, was 0.42 (Figure \ref{fig:gg_gender}).

\subsection{Design of Interactions}
We analysed the number and length of interactions with the robot, the number and type of robots used in the studies, the mode of interaction with the robot, and how the robot was controlled.

\subsubsection{Study Design}
\begin{figure}
    \centering
    \includegraphics[width=\textwidth]{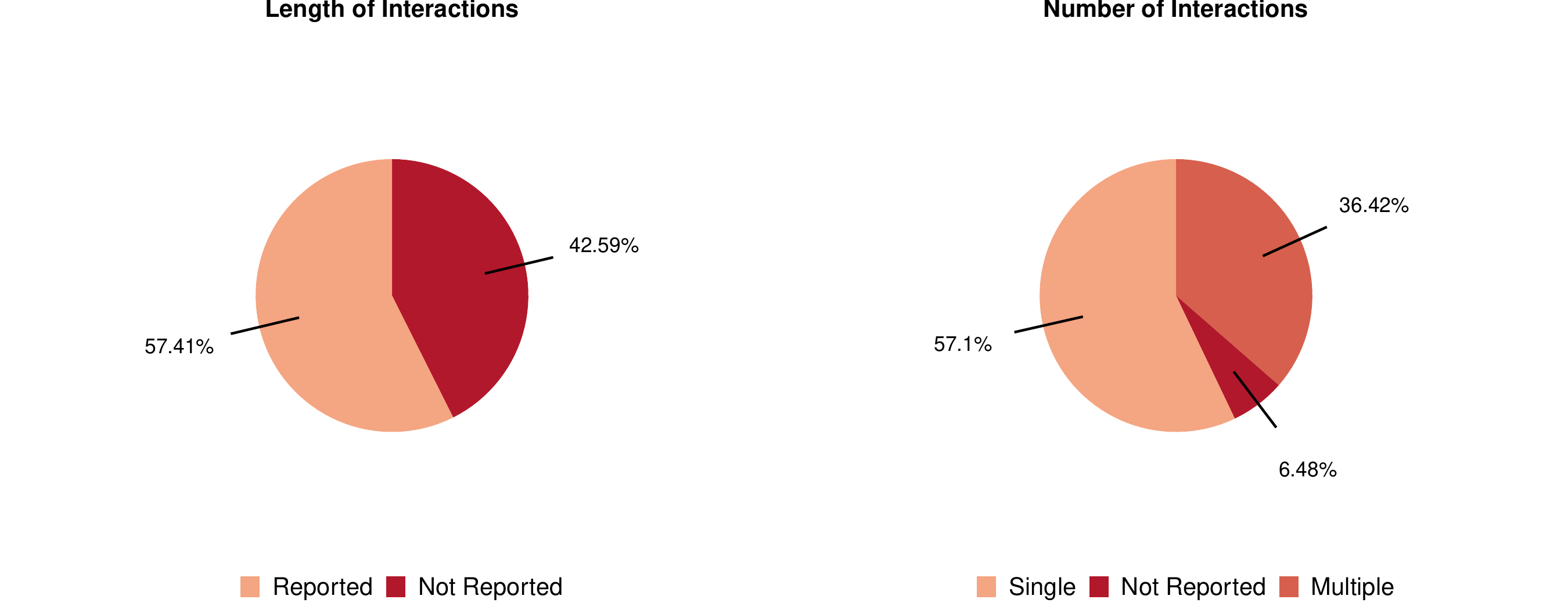}
    \caption{Percentage of studies reporting the number and length of interactions with the robot/s.}
    \label{fig:interaction_pies}
\end{figure}

Regarding the number of interaction sessions, 36.31\% of the studies included in the surveyed papers reported that they conducted studies with multiple sessions (i.e., more than one interaction), and 56.92\% reported that they conducted single-session studies. 6.77\% of the studies in the surveyed manuscripts did not report the number of sessions with the robot.
We then checked whether the studies reported the length of each session with the robot. 42.76\% of the studies included in the manuscripts did not report the length of the interaction, compared to 57.23\% that did (Figure \ref{fig:interaction_pies}).

Many CRI papers have called for more long-term studies to better assess the generalisability of CRI findings. Reporting the details of the interaction, including the number of sessions and the length, therefore allows other researchers to assess the validity and applicability of the study. Furthermore, in the particular cases of educational or healthcare scenarios with children, such information can enable researchers to assess the patterns of engagement or other outcomes that emerge from the interaction. 

In addition, many studies included an ``introduction'' phase with the robot, where the robot introduced itself and/or performed a small activity such as a dance. Reporting such interactions is also important, as it helps understand how children were primed to interact with the robot. This is also relevant from an ethical perspective, as such introductions can help to appropriately calibrate children's expectations about the robot's capabilities, as well as help alleviate any concerns children might have about interacting with the robot. 

\subsubsection{Robot Models}
We extracted the number of robots used in the surveyed papers. 88.62\% of studies reported using one robot, 6.46\% reported using two robots, 2.77\% used three robots, 1.23\% used four robots, 0.62\% reported five robots, while just 1 paper (0.31\%) reported using thirteen robots.

\begin{figure}
    \centering
    \includegraphics[width=0.8\textwidth]{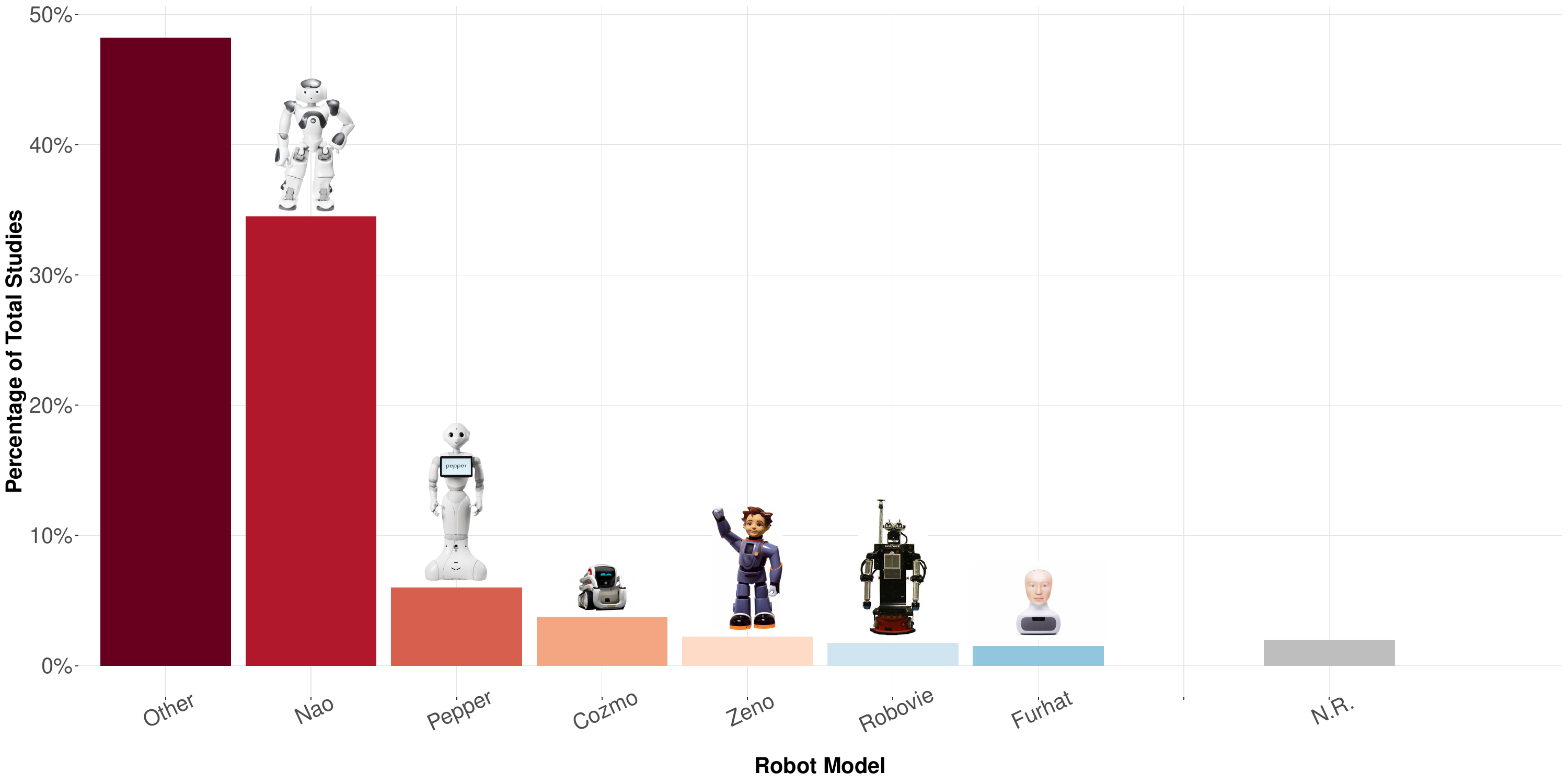}
    \caption{Frequency distribution of the most commonly used robot models in the included papers. Robot models which appeared 5 times or less across the dataset were coded as ``Other''.}
    \label{fig:robot_models}
\end{figure}

The most commonly used robot model was the NAO robot, which was used in 34.50\% of studies, followed by Pepper  (6.00\%), Cozmo (3.75\%), Zeno (2.25\%), Robovie (1.75\%) and Furhat (1.50\%) (see Figure \ref{fig:robot_models}). Jibo, Kaspar, LEGO Mindstorms, Pleo, and QTRobot also appeared 5 times each (1.25\%) across the included studies. The remaining studies (17.50\%) used either other commercial robots (e.g., MiRo-E) or their own custom robot  (see Figure \ref{fig:robot_models} for details).
While only a small percentage (2\%) of the works did not report the robot model that was used, we note that this too is extremely valuable information, as prior work has shown that studies may not replicate if a different robot model is used \cite{irfan2018social}. 

\subsubsection{Mode of interaction}

Only 3 (0.92\%) of the surveyed manuscripts did not report what type of interaction was included in their studies. Of the remaining papers, 89.85\% reported that the interaction was in real-life, 5.23\% reported that the interaction was through video (i.e., participants were asked to watch a video of a robot as in \cite{nijssen2021you} where the authors assessed the effect of anthropomorphic appearance and affective state attributions by watching videos), 1.85\% reported that the interaction was live-mediated (i.e., virtual or augmented reality robot's applications, as in \cite{shahab2022utilizing}, where a social robot in virtual reality was used with children with high functioning autism for music education), 0.92\% reported using a picture of a robot (i.e., children were asked to look at a picture of a robot as in \cite{oranc2020children}, where the authors explored children’s perception of social robots as a source of information across different domains), 0.62\% reported that the interaction was via a virtual robot (not a physical robot), 1 (0.31\%) study reported that the interaction was carried out in augmented reality, and another 1 (0.31\%) reported that the interaction was hybrid (e.g., \cite{yang2020hybrid} combines traditional physical education teaching with a robot teaching).

\begin{figure}
    \centering
    \includegraphics[width=0.8\textwidth]{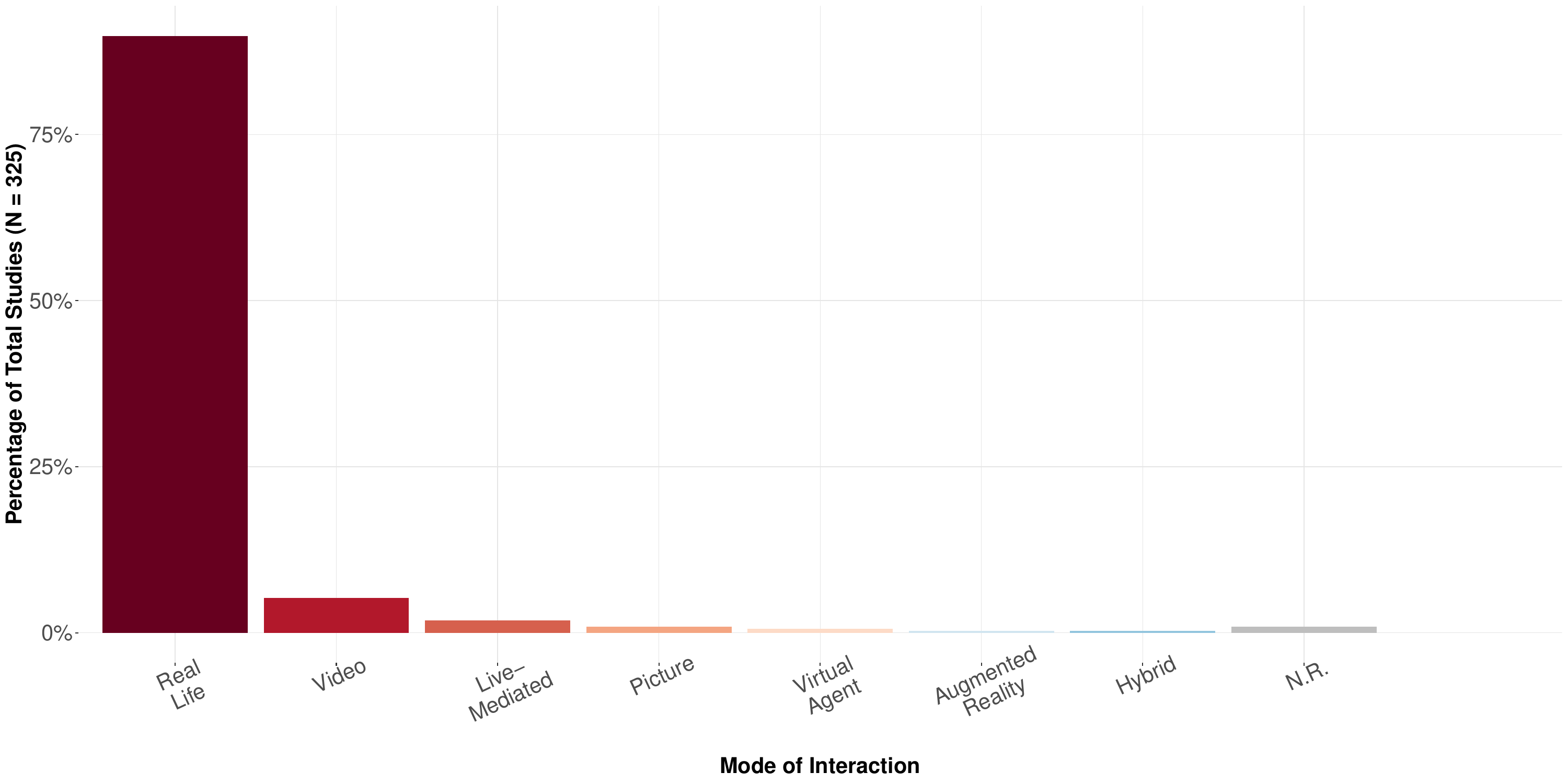}
    \caption{Frequency distribution of the mode of interaction with the robot across the included studies.}
    \label{fig:interaction_mode}
\end{figure}

\subsubsection{Robot operation}

\begin{figure}
    \centering
    \includegraphics[width=0.8\textwidth]{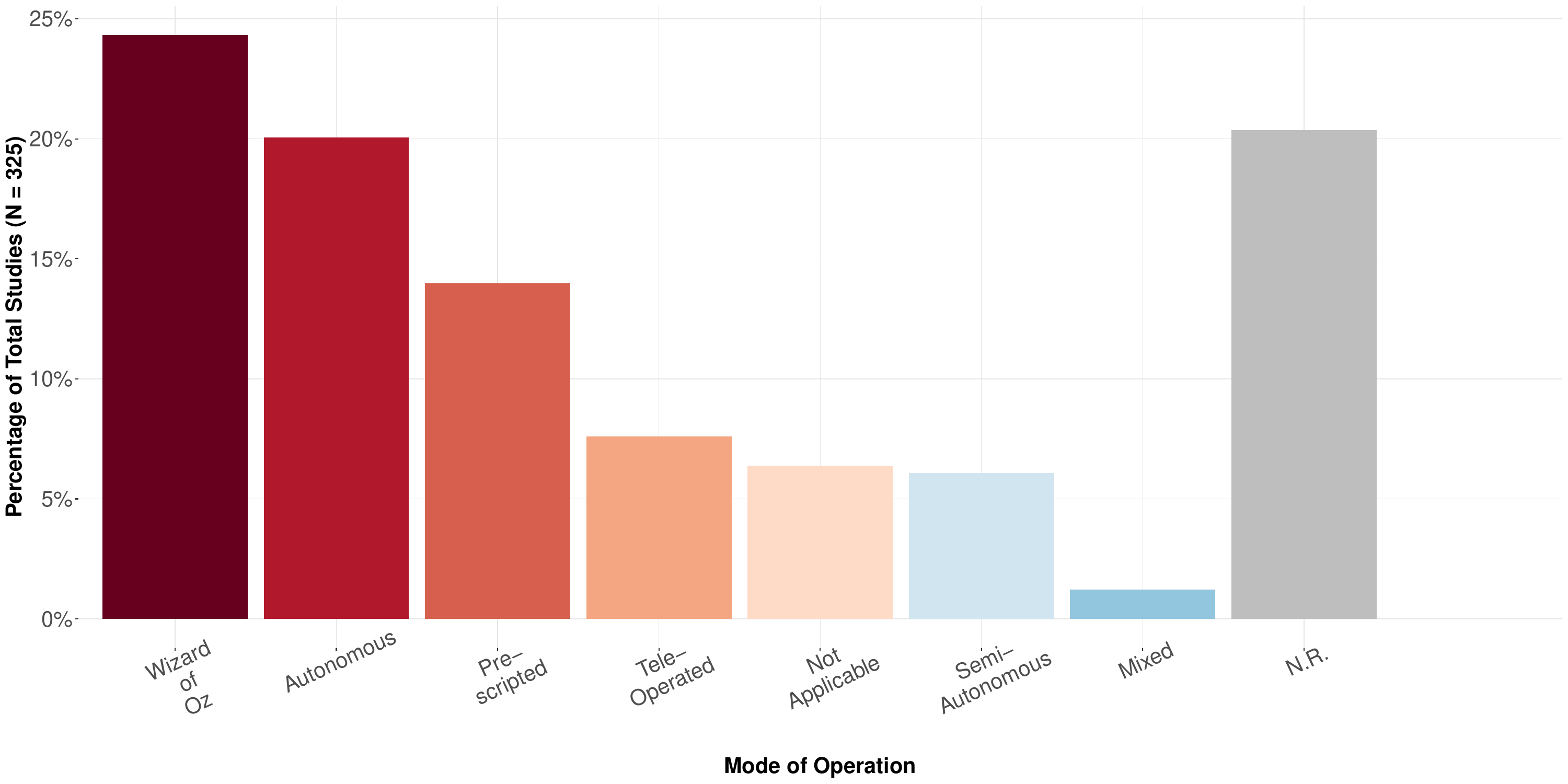}
    \caption{Frequency distribution of the mode of operation of the robot/s across the included studies.}
    \label{fig:robot_operation}
\end{figure}

\begin{figure}
    \centering
    \includegraphics[width=\textwidth]{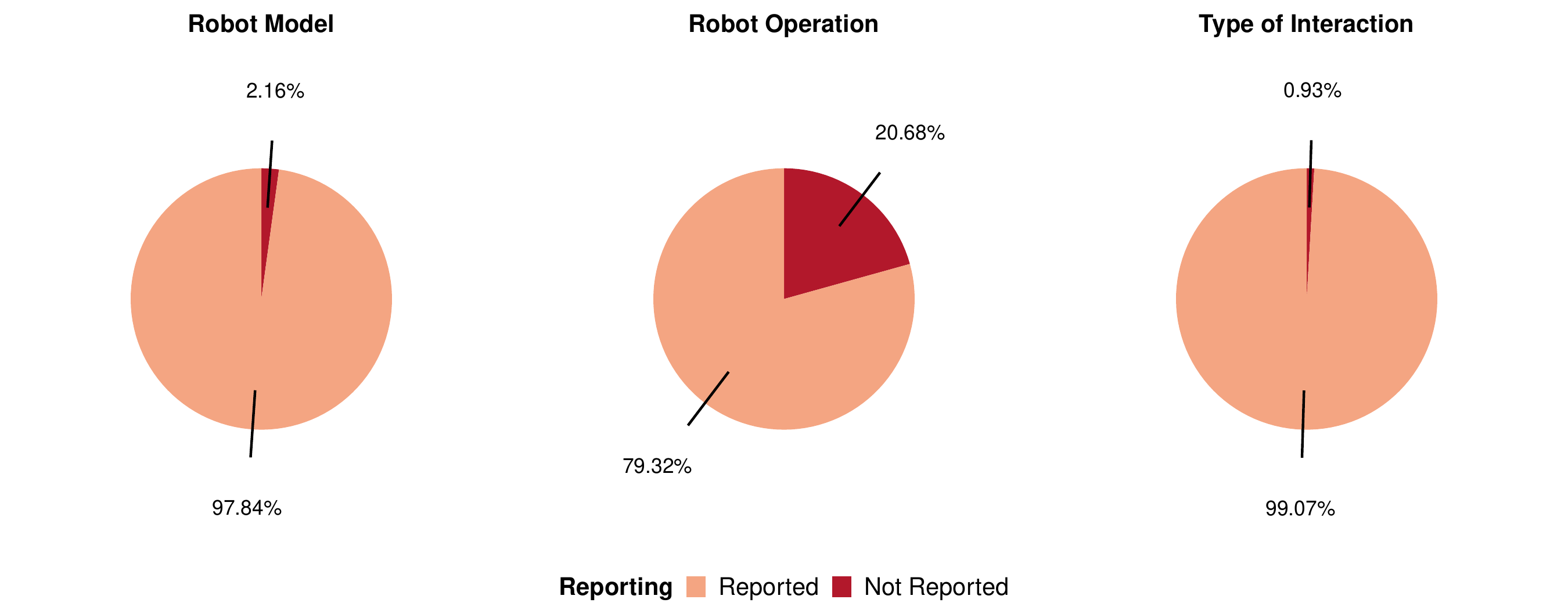}
    \caption{Percentage of studies which reported the robot model, mode of operation, and type of interaction.}
    \label{fig:robot_pies}
\end{figure}

Across all studies, Wizard of Oz was the most commonly used method for operating the robot (24.32\%), followed by autonomous (20.06\%), pre-scripted (13.98\%), teleoperated (7.60\%), and semi-autonomous (6.08\%). 4 studies (1.22\%) reported using more than one method to operate the robot/s, especially if more than one type of robot was involved in the study (e.g., a combination of Wizard of Oz and autonomous, as in \cite{de2022comparing}). 20.32\% of papers did not report how the robot was controlled, and for another 6.38\% the mode of operation was not applicable (for example, if the study did not involve a live interaction with the robot). Figure \ref{fig:robot_operation} displays these findings, and Figure \ref{fig:robot_pies} provides an overview of the reporting on study design across the surveyed papers.

Equally as important, a significant number of the surveyed papers did not report the mode of robot operation (20.92\%). Research in highly interdisciplinary fields can overlook technical contributions, which is why the CRI field might overlook the need to fully describe the systems used. However, this might change if better guidelines and reporting standards are set -- prior work on reproducibility \cite{baxter2016characterising,riek2022wizard} has also pointed out robot operation as an important reporting point towards more reproducible HRI.

\subsection{Ethical and Open Science Guidelines}

Finally, we analysed the adherence of the included papers to current ethical and open science guidelines, including ethical approval and/or informed consent, pre-registration, open access, and availability of code, data, measures, and materials (Figures \ref{fig:ethics_pies} and  \ref{fig:open_pies}).

\subsubsection{Ethical approval and informed consent}

A fundamental aspect of guaranteeing open science is the reporting of the appropriate data protection safeguard measures adopted in compliance with data protection and ethics (e.g., GDPR for EU countries).  58.11\% of papers reported that their studies were approved by a relevant ethics committee (e.g., Institutional Review Board, IRB). 41.22\% of the manuscripts did not report any ethics information, while 2 papers (0.68\%) stated that this was not applicable (e.g., \cite{serholt2020trouble} was conducted in Sweden where ethics approval is not required or mandatory for user studies). 
Analogously, we also checked the papers that reported whether they asked participants to fill out and sign informed consent. 68.92\% of the manuscripts reported asking participants (parents/guardians) to sign an informed consent form, while 31.08\% did not include any consent information. Of the 205 manuscripts that reported asking for informed consent, 49 did not include a corresponding ethical approval. 
Reporting ethical approval and consent is important, not only for readers and reviewers evaluating these studies but also to promote good practices across the field. 

\subsubsection{Preregistration and Open Access}

Only 3.38\% of the 296 manuscripts included in this survey reported that they pre-registered their works (e.g., using the Open Science Framework), while the majority (96.62\%) did not. Over 296 papers, 128 (43.24\%) were open access, while 168 (56.76\%) were not. While most institutions guarantee access to a large number of literature resources, this might still constitute a barrier for researchers to conduct informed research studies.

The low number of pre-registered works could be partially explained by a lack of incentive, as many conferences and journals in HRI do not enforce such practices as a prerequisite for publication. Additionally, preregistration is sometimes met with resistance within the HRI community \cite{gunes2022reproducibility}, where some researchers may feel that they will be unfairly penalised for conducting exploratory analyses with pre-registered study designs. Given this and the pertinence of piloting studies, maybe preregistration could be reviewed as a practice in terms of where it chronologically lays on a study development process -- for instance, to be considered as a stage that takes place after pilot studies, once the study design is finalized.

\begin{figure}
    \centering
    \includegraphics[width=\textwidth]{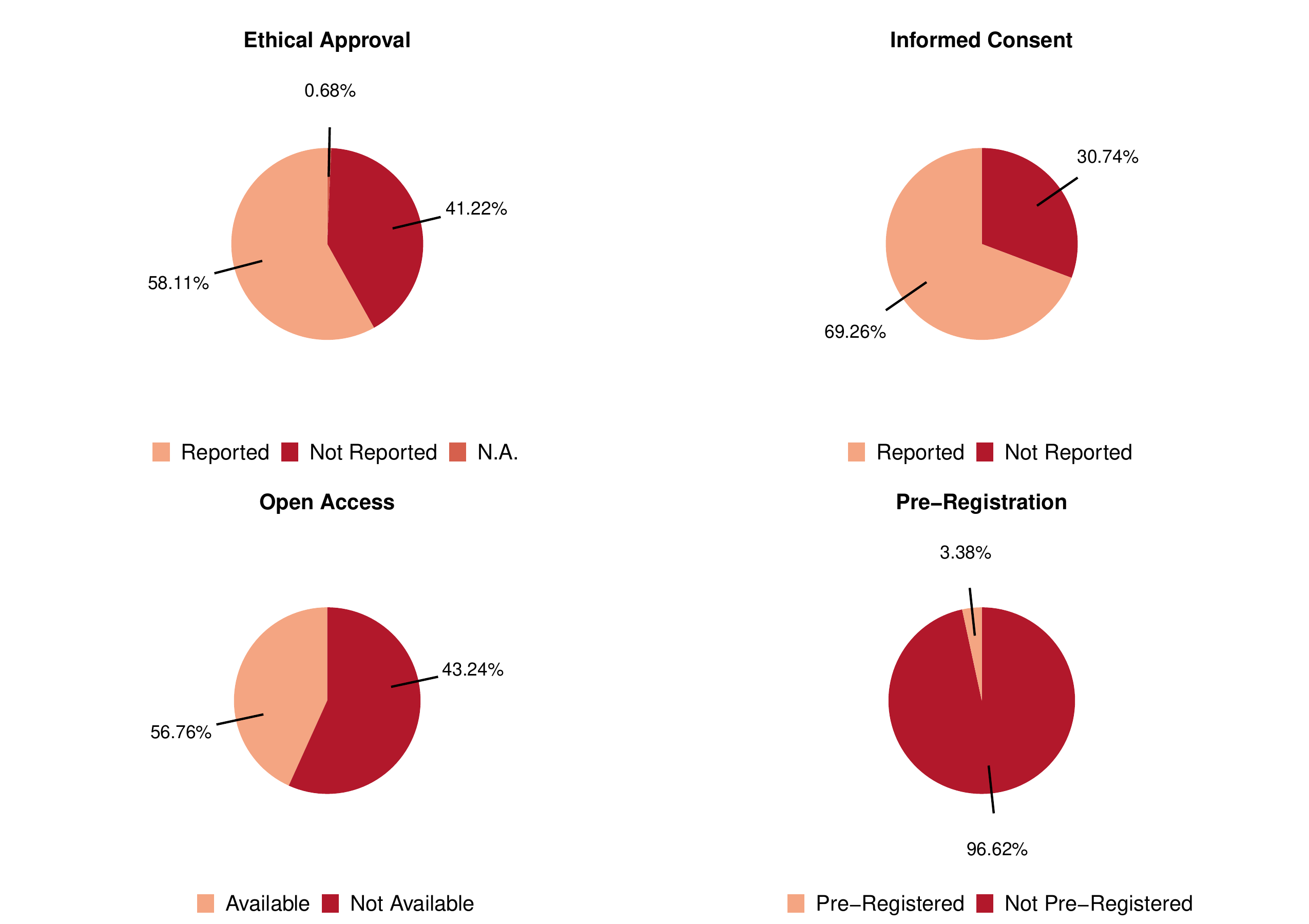}
    \caption{Percentage of studies which reported ethical approval, informed consent, open access, and pre-registration.}
    \label{fig:ethics_pies}
\end{figure}

\subsubsection{Open science}

\begin{figure}
    \centering
    \includegraphics[width=\textwidth]{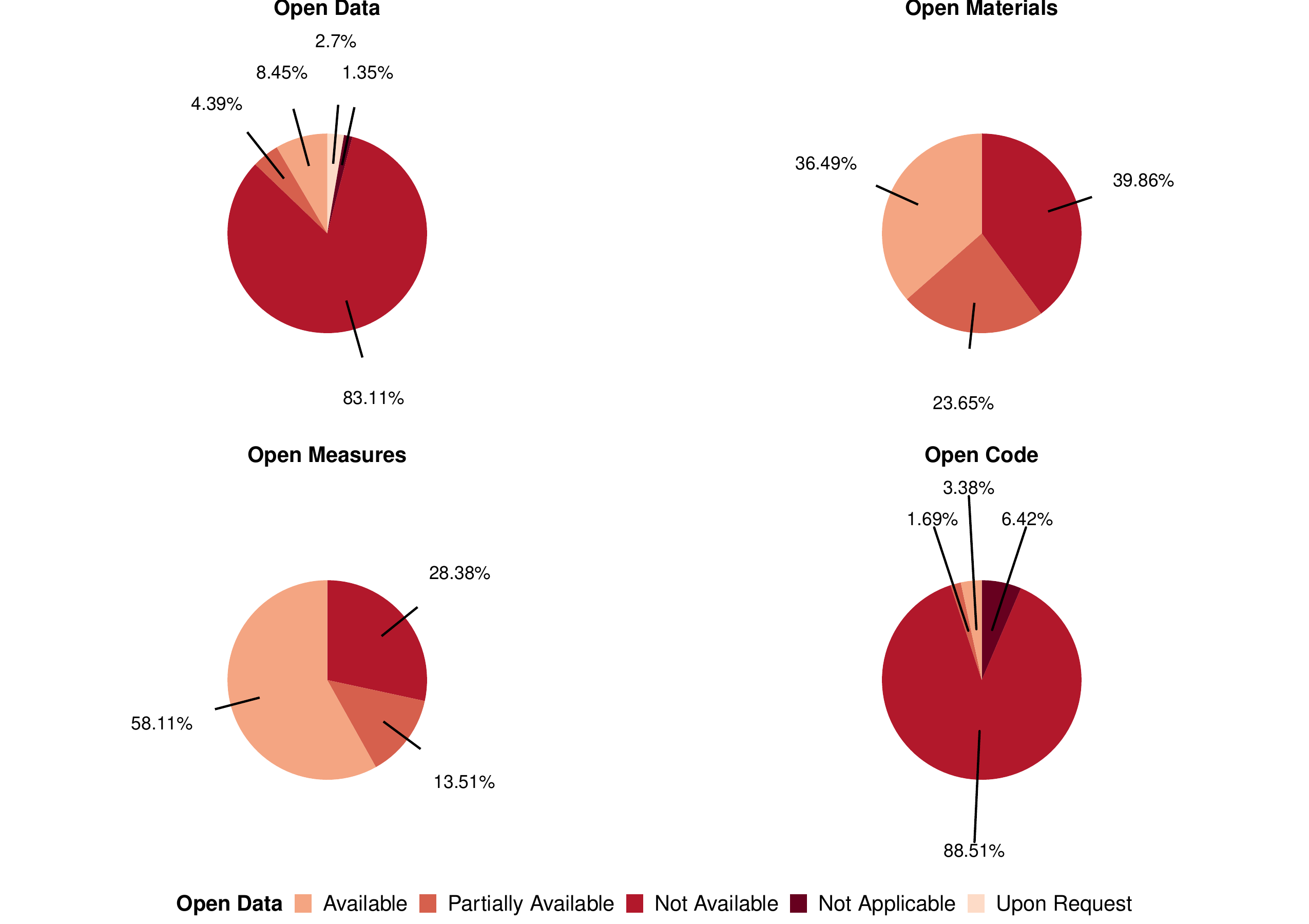}
    \caption{Percentage of studies which had open materials, open measures, open code, and open data.}
    \label{fig:open_pies}
\end{figure}

We then extracted the main open science variables following the European guidelines\footnote{\url{https://research-and-innovation.ec.europa.eu/strategy/strategy-2020-2024/our-digital-future/open-science_en}}: open materials, open measures, open code, and open data (Fig. \ref{fig:open_pies}).
Our results showed that 36.49\% of the included manuscripts reported the materials used in their studies, 23.65\% partially reported the materials used (i.e., the paper reported the exercises delivered with instructions without providing the images of the exercises necessary to accomplish it, e.g., \cite{yang2020hybrid}), 39.86\% did not report any detail of the materials used.

Among the surveyed manuscripts, 58.11\%, reported the measures used for running their analysis and computing their results, 13.51\% partially reported them (e.g., \cite{tian2020understanding}), whereas 28.38\% did not report the measures they used. 
Also, our results showed that only 4.39\% of the surveyed manuscripts provided the code, 1.35\% provided the code partially (e.g., \cite{zhexenova2020comparison}), while the majority of the surveyed papers, 88.51\%, did not provide the code used in their studies. For 17 papers (5.74\%), the code was not applicable because their studies did not include any implementation (e.g., \cite{kewalramani2022multimodal}). 

We also checked whether a code repository was included in the surveyed manuscripts. 4.41\% included a link to a code repository, 1 (0.34\%) partially included this information (i.e., the code link was included but not longer active, e.g., \cite{konijn2022social}). Most papers, 69.15\%, did not include any repository. 

Most of the surveyed papers (83.11\%) did not provide (access to) the data collected or used in the study. 8.45\% provided full access to the data (either in text, in supplementary materials, or an external data file/repository), whereas 4.39\% only provided partial data (e.g.,\cite{tolksdorf2021comparing}). 2.70\% among the surveyed manuscripts reported that the data could be made available upon request, while for 4 papers (1.35\%), this information was not applicable (for example, the study used data that could not be made available due to ethical or privacy constraints).

We also checked how many manuscripts included a study repository or other supplemental materials. We found that most of the studies (224, 75.38\%) did not report any study repository in their manuscript. For 21 manuscripts (7.09\%), the study repository reporting was not applicable (i.e., they haven't implemented any code in their studies). 6.78\% of the manuscripts reported an active study repository, 5.07\% added a statement that the data can be made available upon request, 3.72\% included supplementary materials, 1.69\% reported an inactive link to the study repository.

A very significant portion of the works included in our analysis did not have open materials, i.e. did not provide the different components used for the study (survey questions, experiment code, study data, analysis code, etc. -- for instance, only 4.39\% of papers provided code). These are important elements of a study that are prominent for not only thorough evaluation of the study methods but also for reproducibility and replicability. We note, as \citet{baxter2016characterising} mention, that CRI poses particular challenges when it comes to data sharing, namely identifiable data. Children's identity protection and right to privacy should be a top priority for all researchers, and thus any reproducibility good practices must never override these considerations. Nonetheless, strategies for bettering the sharing of study materials should be discussed, developed, and implemented whenever possible. The lack of guidelines in reporting such data in turn posited challenges and limitations in extracting the variables described in Data Extraction and Analysis section.

\section{Guidelines to improve reproducibilty}

Child-robot interaction tackles similar obstacles to replicate published studies as its parent field, human-robot interaction \cite{baxter2016characterising}. However, CRI is a more recent field, which thus suffers from the problem of a lack of definitions and standardised guidelines. To add to this, the interdisciplinarity of the field results in a multitude of not only application contexts but also author disciplines. Different disciplines value different styles of reporting, thus enhancing the inconsistency we observe in reporting in CRI works. 
Hence, we conducted this systematic review and the findings of this work enable us to distill a set of guidelines to improve reproducibility as follows.

Our results suggest that most of the surveyed papers reported investigating applications related to health (e.g., \cite{neerincx2021social, beyer2020effects})  and education (e.g., \cite{lopez2020using, de2022comparing})  for the target group of CRI, i.e., children, as depicted in Figure \ref{fig:applications}. In those application scenarios, literature \cite{mott2022robot} highlights the importance of ethical considerations for such a population due to concerns about trust and deception, data privacy and security. \citet{langer2023ethical} also pinpoints the necessity of ensuring transparency in consent data usage, limitations, and destruction \cite{livingstone2018european}. However, the surveyed papers still lack a comprehensive reporting of ethical approval, informed consent and preregistration as displayed in Figure \ref{fig:ethics_pies}. 
Therefore, we recommend \textit{reporting the ethical approval, and when possible, preregistration of studies}. Detailed reporting of how consent was acquired (for example, by providing the consent form in a study repository) is also pivotal.

We found that most of the surveyed papers (e.g., \cite{ismail2021analysis, calvo2021effects}) reported details about the system and the study design, but most of them had small sample sizes (with a median of 25 participants), as shown in Figure \ref{fig:sample_sizes}. Thus, replication studies are called for. In order to replicate a study, however, researchers require an in-depth description of the system used (robot model and mode of operation, the deployment scenario --  e.g., what did the room look like, who was in the room).
In line with application settings and vulnerable population groups, past works \cite{dipaola2023fundamental,hoffman2020} suggested that it is not enough to briefly describe the system and study design but their reporting must be thoughtful. \citet{fraune2022lessons} also presented lessons learned for designing HRI studies by highlighting that researchers should properly document the task and the entire experiment in their papers so others can replicate the study with the same conditions and a similar environment.
Therefore, we recommend that \textit{robotic systems and study design should be described in detail}. Ideally, code for robot operation and any other aspects of the system should be freely available for any researcher attempting to replicate the study.

We found that not all the surveyed papers reported information about the demographics of children included in their studies (see Figure \ref{fig:demographics}). 
Literature \cite{cordero2022reporting, fraune2022lessons} highlights the importance of thorough reporting on demographics (age, gender, and inclusion/exclusion criteria). 
For example, \citet{fraune2022lessons} reported that children in different developmental stages may display different behaviour when interacting with robots. Another consideration is the fact that the range of age is extremely relevant in children: when designing an CRI study, a 2-year age range could be too broad, due to differences in cognitive developmental stages.
Information about demographics allows researchers to understand the results within the child developmental context but also provides the groundwork for comparing replication studies (and why, sometimes, results might not replicate \cite{ullman2021challenges,leichtmann2022crisis}). We then recommend that \textit{recruitment methods and exclusion criteria should also be provided.}

Our results show that the surveyed papers lacked in providing materials, measures and especially open code and data, as depicted in Figure \ref{fig:open_pies}. 
As mentioned, past works \cite{koole2012replications, stower2021interdisciplinary} stressed that providing study materials is paramount for achieving reporting transparency, promoting reproducibility, and incentivising replicability. Thus, we recommend \textit{researchers maintain study repositories and provide as many resources as possible (deployment and analysis code, surveys, forms, etc.).}

Following our findings in Section \ref{sec:results} and similarly to prior work \cite{gunes2022reproducibility, baxter2016characterising}, in Table \ref{tab:checklist} we provide a checklist that authors can use to self-assess the quality of their reporting. We note that manuscript length restrictions may exist, for which other strategies can be used (supplementary material, online repository, etc.). Publishing venues can exclude these sections from the page limit; in any case, this information should be easily traceable for readers.

\renewcommand{\arraystretch}{1.1}
\begin{table}[htb!]
\caption{Checklist to systematically report child-robot interaction studies in particular, that can be also extended to user studies with robots in general.}
    \centering
     \resizebox{0.75\textwidth}{!}
   {
    \begin{tabular}{|l|l|l|}
     \hline
    \textbf{Study Stage}     & \textbf{List of items} & \textbf{We recommend reporting..}\\
    \hline
    \textbf{Study}
        &$\square$ Preregistration & .. if you have preregistered the study (before or after piloting)\\
        \textbf{Conceptualisation} &$\square$ Ethical approval & .. whether the study was approved by an ethics committee\\
        &$\square$ Sample Size & .. the target number of participants aiming to be recruited, \\
        && and how this number was obtained (e.g, power-analysis, \\
        && pilot studies, previous work)\\
    \hline
    \textbf{System Design}
        &$\square$ Robot Model & .. which robot platform was used in the study (e.g., Nao)\\
        &$\square$ Mode of interaction & .. the modality of the interaction (e.g., real-life, video) \\
        && between the human and the robot\\
        &$\square$ Robot operation & .. the autonomy level of the robot, both in terms of \\
        && perception and generation (e.g., autonomous, wizard-of-Oz)\\
        &$\square$ Open code & .. the code used for implementing the interaction with \\
        && the robot, specifying the study repository if applicable. \\
    \hline
    \textbf{Study Design}   
        &$\square$ Pilot tests & .. if applicable, how pilot studies were carried out\\
        \textbf{and Deployment}&& and results/conclusions from these studies \\
        &$\square$ Consent & .. how informed consent was provided (ideally, share the \\
        && consent form used for the study)\\
        &$\square$ Recruitment criteria & .. which criteria and methods were used for recruitment \\
        &$\square$ Population & .. the population involved in the study\\
        && (neurotypical, neurodivergent, etc.) \\
        &$\square$ Number of participants & .. the number of participants involved in the study  \\
        &$\square$ Age & .. the age range and mean age of the participants \\
        && (if applicable, provide this for each condition of study)\\
        &$\square$ Gender  & .. the genders of participants involved in the study \\
        && (if applicable, provide this for each condition of study)\\
        &$\square$ Screening information & .. whether the participants were screened and the \\
        && information about age and gender before and after\\
        && the screening exclusion\\
        &$\square$ Experiment setting & .. what was the setting when running the study \\
        && (what the room looked like, who was present, etc.) \\
        &$\square$ Number of sessions & .. the number of sessions during which the participants \\
        && interacted with the robot\\
        &$\square$ Length of interaction(s) & .. the length (i.e., duration) of each interaction with the robot\\
        &$\square$ Debriefing & .. how participants were debriefed about\\
        && study goals and research questions \\
        &$\square$ Open materials and measures & .. the measures used for collecting the data \\
        && (surveys, task materials, interview questions, etc.)\\
    \hline
    \textbf{Analysis}  
        &$\square$ Open analysis & .. the analysis conducted for analysing the data collected \\
        && (if applicable, provide script used) \\
        &$\square$ Open data & .. the data collected if fully anonymised and \\
        && in compliance with privacy rights, or the reason why \\
        && these data cannot be openly available \\

    \hline
    \end{tabular}}
    \label{tab:checklist}
\end{table}

\section{General Discussion}
\label{sec:discussion}

This section describes from a high-level the implications of our review in terms of reporting, transparency, and reproducibility, and the open challenges and future directions for reproducibility in CRI.

\subsection{Reporting, Transparency, and Reproducibility}

Challenges specific to reproducibility in CRI are found in our analysis. Namely, studies in CRI often involve vulnerable population demographics (non-neurotypical groups), and there is a tendency towards high-stakes scenarios. This requires additional considerations when designing a study. The study's motivation and design need to be carefully planned and controlled, and participants and their guardians need to be adequately informed about all the foreseeable risks and benefits of their participation. Reporting ethical approval and consent is important, not only for readers and reviewers evaluating these studies but also to promote good practices across the field \cite{tolksdorf2021ethical}. Further, with vulnerable populations and high-stakes application scenarios, the need for literature that appropriately reports on the recruitment process and challenges, as well as provides a clear description of the study design and procedure, becomes obvious.

The interdisciplinarity of researchers and research questions in CRI is also an emergent finding from this work. This is reflected in inconsistent reporting, which constituted a challenge even for the data extraction method used in this analysis. However, the lack of standardised methods and metrics, and interdisciplinary collaborations can be seen as not a limitation but rather as an opportunity to build theory and good practices in the field, that can be communicated and applied effectively \cite{chrysostomou2017standard,stower2023workshop}.

While reporting is one of the primary elements for promoting reproducibility, we recognise that other constraints may arise in realistic settings. The lack of standardised questionnaires and metrics can be potentially addressed by Institutional Review Boards (IRBs) or ethics review committees of the institutions of the authors, although this might be impractical in interdisciplinary collaborations (as is often the case with CRI), as the guidelines for surveys can conflict. Another factor hindering extensive reporting is also manuscript length restrictions, which are included in the paper submission guidelines for most conferences and other publishing venues. However, we identify strategies that can help address these restrictions, namely including these data as supplementary material, or creating (and actively maintaining) study repositories. Below, we reflect on the implementation of better practices, from the perspective of different stakeholders.

\subsection{Open Challenges and Future Directions}

Enforcing appropriate reporting in papers to enable reproducibility in the field of CRI requires a collective effort involving various stakeholders. While each entity plays a role, it is essential to recognise the shared responsibility among publishing venues, institutions, and authors \cite{koole2012replications, munafo2017manifesto}.

Publishing venues hold a significant responsibility in ensuring the integrity and reproducibility of research. They can establish and enforce guidelines that promote transparent and rigorous reporting practices. Journals and conference proceedings can require authors to provide detailed methodological information, access to datasets, and source code when applicable. Additionally, they can encourage replication studies and require authors to address limitations and potential sources of variation explicitly. We provide the example of the ACM Interaction Design and Children conference (IDC) \cite{IDC}, which requires authors to submit papers with a ``Selection and Participation of Children'' section in which \textit{``the authors of the paper should describe how children were selected (...), what consent processes were followed (...), how they were treated, how data sharing was communicated, and especially any additional ethical considerations''}. Analogously the Affective Computing and Intelligence Interaction conference (ACII) has introduced a mandatory section on ``Ethical Statement'' since 2022 because \textit{``a large diversity of research done in the Affective Computing community today also means that there is a large diversity of ethical issues. Not every project will encounter every issue, but almost all projects will encounter some issues"}. 

Institutions also play a vital role in fostering reproducibility. They can promote research integrity and provide support and resources for researchers to conduct reproducible studies. Institutions can offer training and workshops on transparent research practices, data management, and statistical analysis. They can also encourage collaborative efforts among researchers to replicate and validate findings. Furthermore, institutions can establish policies that incentivise and reward researchers for practicing open science and reproducibility, such as recognising open access publications or replication studies in evaluation and promotion processes.

Ethical approval procedures to conduct studies with human participants or children are in place in the majority of countries and institutions. 
While each country, university, or department has its own procedures and requirements, adhering to these standards and reporting them can not only provide other researchers with the knowledge that the studies were ethically approved but also removes barriers to reproducing the study, given that it adheres with the ethical procedures in other countries. In the case of vulnerable populations, particularly children with any type of difficulties (e.g. autism or dysgraphia) being able to reproduce the studies goes beyond notions of reproducibility to the realms of helping them through utilising recent scientific advances. 

Finally, authors themselves hold direct responsibility for reporting their research in a manner that allows for reproducibility. We should strive to provide clear and comprehensive descriptions of their methods, materials, and data collection procedures. Authors can adhere to reporting guidelines specific to CRI or follow broader guidelines like the Transparency and Openness Promotion (TOP) guidelines \footnote{\url{https://www.cos.io/initiatives/top-guidelines}}. Additionally, where possible, authors should openly share their datasets, code, and any other relevant research materials to enable independent verification and replication of their work.

\section{Conclusion}
\label{sec:conclusion}

Ensuring appropriate reporting and reproducibility in science requires a collective effort. 
The current review investigates how the field of child-robot interaction fares on different reproducibility metrics with an additional overview of the field by analysing 225 CRI papers published from 2020-2022. 
Our findings revealed that considering the targeted population (i.e., children) and high-stakes scenarios  (e.g. healthcare), the field has significant progress yet to make. 
We identified different areas where reporting was not comprehensive enough to enable reproducible work. Based on these, we developed a set of guidelines that could help researchers in the field to reflect on various stages of their study design, development, and reporting to ensure contribution to the reproducibility culture. 
Publishing venues, institutions, and authors all play crucial roles in establishing and promoting standards and practices that support transparent and replicable research. 
By working together, these stakeholders can contribute to advancing the field of CRI and fostering a scientific community committed to rigorous and reliable research practices.

\section*{Acknowledgments}
This work was partially funded by the S-FACTOR project from NordForsk, Digital Futures Research Center, and Vinnova Competence Center for Trustworthy Edge Computing Systems and Applications at KTH.  M.S. and H.G. are supported by the EPSRC/UKRI under grant ref. EP/R030782/1 (ARoEQ).
We thank Dr. Mart Couto and Dr. Shruti Chandra for sharing the list of their screened paper on child-robot interaction from January 2020 to December.
We would also like to thank the organisers of the RPL Summer School, for facilitating the initial conception of this work.

\section*{Author Contributions}
M.S. and R.S. conceived of the initial study. M.S., R.S, E.Y, M.P, and N.A., performed paper screening and data extraction. R.S. analysed the data. M.S. and R.S. wrote the methods and results. M.S. wrote the guidelines and checklist for future CRI researchers. E.Y. and M.P. wrote the introduction and general discussion sections.  H.G. and I.L. oversaw the project and provided guidance. All authors proofread the manuscript. 

\bibliographystyle{ACM-Reference-Format}
\balance
\bibliography{references.bib}


\begin{thebibliography}{64}


\ifx \showCODEN    \undefined \def \showCODEN     #1{\unskip}     \fi
\ifx \showDOI      \undefined \def \showDOI       #1{#1}\fi
\ifx \showISBNx    \undefined \def \showISBNx     #1{\unskip}     \fi
\ifx \showISBNxiii \undefined \def \showISBNxiii  #1{\unskip}     \fi
\ifx \showISSN     \undefined \def \showISSN      #1{\unskip}     \fi
\ifx \showLCCN     \undefined \def \showLCCN      #1{\unskip}     \fi
\ifx \shownote     \undefined \def \shownote      #1{#1}          \fi
\ifx \showarticletitle \undefined \def \showarticletitle #1{#1}   \fi
\ifx \showURL      \undefined \def \showURL       {\relax}        \fi
\providecommand\bibfield[2]{#2}
\providecommand\bibinfo[2]{#2}
\providecommand\natexlab[1]{#1}
\providecommand\showeprint[2][]{arXiv:#2}

\bibitem[\protect\citeauthoryear{??}{IDC}{[n.d.]}]%
        {IDC}
 \bibinfo{year}{[n.d.]}\natexlab{}.
\newblock \bibinfo{title}{ACM Interaction Design and Children 2023}.
\newblock
  \bibinfo{howpublished}{\url{https://idc.acm.org/2023/full-short-papers/}}.
\newblock
\newblock
\shownote{Accessed: 2023-05-17.}


\bibitem[\protect\citeauthoryear{Abbasi, Spitale, Anderson, Ford, Jones, and
  Gunes}{Abbasi et~al\mbox{.}}{2022a}]%
        {abbasi2022can}
\bibfield{author}{\bibinfo{person}{Nida~Itrat Abbasi}, \bibinfo{person}{Micol
  Spitale}, \bibinfo{person}{Joanna Anderson}, \bibinfo{person}{Tamsin Ford},
  \bibinfo{person}{Peter~B Jones}, {and} \bibinfo{person}{Hatice Gunes}.}
  \bibinfo{year}{2022}\natexlab{a}.
\newblock \showarticletitle{Can robots help in the evaluation of mental
  wellbeing in children? An empirical study}. In \bibinfo{booktitle}{\emph{2022
  31st IEEE International Conference on Robot and Human Interactive
  Communication (RO-MAN)}}. IEEE, \bibinfo{pages}{1459--1466}.
\newblock


\bibitem[\protect\citeauthoryear{Abbasi, Spitale, Jones, and Gunes}{Abbasi
  et~al\mbox{.}}{2022b}]%
        {abbasi2022measuring}
\bibfield{author}{\bibinfo{person}{Nida~Itrat Abbasi}, \bibinfo{person}{Micol
  Spitale}, \bibinfo{person}{Peter~B Jones}, {and} \bibinfo{person}{Hatice
  Gunes}.} \bibinfo{year}{2022}\natexlab{b}.
\newblock \showarticletitle{Measuring mental wellbeing of children via
  human-robot interaction: Challenges and opportunities}.
\newblock \bibinfo{journal}{\emph{Interaction Studies}} \bibinfo{volume}{23},
  \bibinfo{number}{2} (\bibinfo{year}{2022}), \bibinfo{pages}{157--203}.
\newblock


\bibitem[\protect\citeauthoryear{Bainbridge, Hart, Kim, and
  Scassellati}{Bainbridge et~al\mbox{.}}{2011}]%
        {bainbridge2011benefits}
\bibfield{author}{\bibinfo{person}{Wilma~A Bainbridge},
  \bibinfo{person}{Justin~W Hart}, \bibinfo{person}{Elizabeth~S Kim}, {and}
  \bibinfo{person}{Brian Scassellati}.} \bibinfo{year}{2011}\natexlab{}.
\newblock \showarticletitle{The benefits of interactions with physically
  present robots over video-displayed agents}.
\newblock \bibinfo{journal}{\emph{International Journal of Social Robotics}}
  \bibinfo{volume}{3} (\bibinfo{year}{2011}), \bibinfo{pages}{41--52}.
\newblock


\bibitem[\protect\citeauthoryear{Baxter, Kennedy, Senft, Lemaignan, and
  Belpaeme}{Baxter et~al\mbox{.}}{2016}]%
        {baxter2016characterising}
\bibfield{author}{\bibinfo{person}{Paul Baxter}, \bibinfo{person}{James
  Kennedy}, \bibinfo{person}{Emmanuel Senft}, \bibinfo{person}{Severin
  Lemaignan}, {and} \bibinfo{person}{Tony Belpaeme}.}
  \bibinfo{year}{2016}\natexlab{}.
\newblock \showarticletitle{From characterising three years of HRI to
  methodology and reporting recommendations}. In \bibinfo{booktitle}{\emph{2016
  11th ACM/IEEE International Conference on Human-Robot Interaction (HRI)}}.
  IEEE, \bibinfo{pages}{391--398}.
\newblock


\bibitem[\protect\citeauthoryear{Belpaeme, Baxter, De~Greeff, Kennedy, Read,
  Looije, Neerincx, Baroni, and Zelati}{Belpaeme et~al\mbox{.}}{2013}]%
        {belpaeme2013child}
\bibfield{author}{\bibinfo{person}{Tony Belpaeme}, \bibinfo{person}{Paul
  Baxter}, \bibinfo{person}{Joachim De~Greeff}, \bibinfo{person}{James
  Kennedy}, \bibinfo{person}{Robin Read}, \bibinfo{person}{Rosemarijn Looije},
  \bibinfo{person}{Mark Neerincx}, \bibinfo{person}{Ilaria Baroni}, {and}
  \bibinfo{person}{Mattia~Coti Zelati}.} \bibinfo{year}{2013}\natexlab{}.
\newblock \showarticletitle{Child-robot interaction: Perspectives and
  challenges}. In \bibinfo{booktitle}{\emph{Social Robotics: 5th International
  Conference, ICSR 2013, Bristol, UK, October 27-29, 2013, Proceedings 5}}.
  Springer, \bibinfo{pages}{452--459}.
\newblock


\bibitem[\protect\citeauthoryear{Beran, Ramirez-Serrano, Kuzyk, Fior, and
  Nugent}{Beran et~al\mbox{.}}{2011}]%
        {beran2011understanding}
\bibfield{author}{\bibinfo{person}{Tanya~N Beran}, \bibinfo{person}{Alejandro
  Ramirez-Serrano}, \bibinfo{person}{Roman Kuzyk}, \bibinfo{person}{Meghann
  Fior}, {and} \bibinfo{person}{Sarah Nugent}.}
  \bibinfo{year}{2011}\natexlab{}.
\newblock \showarticletitle{Understanding how children understand robots:
  Perceived animism in child--robot interaction}.
\newblock \bibinfo{journal}{\emph{International Journal of Human-Computer
  Studies}} \bibinfo{volume}{69}, \bibinfo{number}{7-8} (\bibinfo{year}{2011}),
  \bibinfo{pages}{539--550}.
\newblock


\bibitem[\protect\citeauthoryear{Beyer-Wunsch and Reichstein}{Beyer-Wunsch and
  Reichstein}{2020}]%
        {beyer2020effects}
\bibfield{author}{\bibinfo{person}{Pia Beyer-Wunsch} {and}
  \bibinfo{person}{Christopher Reichstein}.} \bibinfo{year}{2020}\natexlab{}.
\newblock \showarticletitle{Effects of a humanoid robot on the well-being for
  hospitalized children in the pediatric clinic-An experimental study}.
\newblock \bibinfo{journal}{\emph{Procedia Computer Science}}
  \bibinfo{volume}{176} (\bibinfo{year}{2020}), \bibinfo{pages}{2077--2087}.
\newblock


\bibitem[\protect\citeauthoryear{Button, Ioannidis, Mokrysz, Nosek, Flint,
  Robinson, and Munaf{\`o}}{Button et~al\mbox{.}}{2013}]%
        {button2013power}
\bibfield{author}{\bibinfo{person}{Katherine~S Button},
  \bibinfo{person}{John~PA Ioannidis}, \bibinfo{person}{Claire Mokrysz},
  \bibinfo{person}{Brian~A Nosek}, \bibinfo{person}{Jonathan Flint},
  \bibinfo{person}{Emma~SJ Robinson}, {and} \bibinfo{person}{Marcus~R
  Munaf{\`o}}.} \bibinfo{year}{2013}\natexlab{}.
\newblock \showarticletitle{Power failure: why small sample size undermines the
  reliability of neuroscience}.
\newblock \bibinfo{journal}{\emph{Nature reviews neuroscience}}
  \bibinfo{volume}{14}, \bibinfo{number}{5} (\bibinfo{year}{2013}),
  \bibinfo{pages}{365--376}.
\newblock


\bibitem[\protect\citeauthoryear{Calvo-Barajas, Perugia, and
  Castellano}{Calvo-Barajas et~al\mbox{.}}{2021}]%
        {calvo2021effects}
\bibfield{author}{\bibinfo{person}{Natalia Calvo-Barajas},
  \bibinfo{person}{Giulia Perugia}, {and} \bibinfo{person}{Ginevra
  Castellano}.} \bibinfo{year}{2021}\natexlab{}.
\newblock \showarticletitle{The effects of motivational strategies and goal
  attainment on children’s trust in a virtual social robot: A pilot study}.
  In \bibinfo{booktitle}{\emph{Interaction Design and Children}}.
  \bibinfo{pages}{537--541}.
\newblock


\bibitem[\protect\citeauthoryear{Catania, Spitale, and Garzotto}{Catania
  et~al\mbox{.}}{2022}]%
        {catania2022conversational}
\bibfield{author}{\bibinfo{person}{Fabio Catania}, \bibinfo{person}{Micol
  Spitale}, {and} \bibinfo{person}{Franca Garzotto}.}
  \bibinfo{year}{2022}\natexlab{}.
\newblock \showarticletitle{Conversational agents in therapeutic interventions
  for neurodevelopmental disorders: a survey}.
\newblock \bibinfo{journal}{\emph{ACM Computing Surveys (CSUR)}}
  (\bibinfo{year}{2022}).
\newblock


\bibitem[\protect\citeauthoryear{Chen, Park, Zhang, and Breazeal}{Chen
  et~al\mbox{.}}{2020}]%
        {chen2020impact}
\bibfield{author}{\bibinfo{person}{Huili Chen}, \bibinfo{person}{Hae~Won Park},
  \bibinfo{person}{Xiajie Zhang}, {and} \bibinfo{person}{Cynthia Breazeal}.}
  \bibinfo{year}{2020}\natexlab{}.
\newblock \showarticletitle{Impact of interaction context on the student
  affect-learning relationship in child-robot interaction}. In
  \bibinfo{booktitle}{\emph{Proceedings of the 2020 ACM/IEEE International
  Conference on Human-Robot Interaction}}. \bibinfo{pages}{389--397}.
\newblock


\bibitem[\protect\citeauthoryear{Chrysostomou, Barattini, Kildal, Wang, Fo,
  Dautenhahn, Ferland, Tapus, and Virk}{Chrysostomou et~al\mbox{.}}{2017}]%
        {chrysostomou2017standard}
\bibfield{author}{\bibinfo{person}{Dimitrios Chrysostomou},
  \bibinfo{person}{Paolo Barattini}, \bibinfo{person}{Johan Kildal},
  \bibinfo{person}{Yue Wang}, \bibinfo{person}{Jacopo Fo},
  \bibinfo{person}{Kerstin Dautenhahn}, \bibinfo{person}{Francois Ferland},
  \bibinfo{person}{Adriana Tapus}, {and} \bibinfo{person}{Gurvinder~S. Virk}.}
  \bibinfo{year}{2017}\natexlab{}.
\newblock \showarticletitle{ReHRI'17 - Towards Reproducible HRI Experiments:
  Scientific Endeavors, Benchmarking and Standardization}. In
  \bibinfo{booktitle}{\emph{Proceedings of the Companion of the 2017 ACM/IEEE
  International Conference on Human-Robot Interaction}} (Vienna, Austria)
  \emph{(\bibinfo{series}{HRI '17})}. \bibinfo{publisher}{Association for
  Computing Machinery}, \bibinfo{address}{New York, NY, USA},
  \bibinfo{pages}{421–422}.
\newblock
\showISBNx{9781450348850}
\urldef\tempurl%
\url{https://doi.org/10.1145/3029798.3029800}
\showDOI{\tempurl}


\bibitem[\protect\citeauthoryear{Clabaugh, Jain, Thiagarajan, Shi, Mathur,
  Mahajan, Ragusa, and Matari{\'c}}{Clabaugh et~al\mbox{.}}{2020}]%
        {clabaugh2020month}
\bibfield{author}{\bibinfo{person}{Caitlyn Clabaugh}, \bibinfo{person}{Shomik
  Jain}, \bibinfo{person}{Balasubramanian Thiagarajan},
  \bibinfo{person}{Zhonghao Shi}, \bibinfo{person}{Leena Mathur},
  \bibinfo{person}{Kartik Mahajan}, \bibinfo{person}{Gisele Ragusa}, {and}
  \bibinfo{person}{Maja Matari{\'c}}.} \bibinfo{year}{2020}\natexlab{}.
\newblock \showarticletitle{Month-long, in-home socially assistive robot for
  children with diverse needs}. In \bibinfo{booktitle}{\emph{Proceedings of the
  2018 International Symposium on Experimental Robotics}}. Springer,
  \bibinfo{pages}{608--618}.
\newblock


\bibitem[\protect\citeauthoryear{Collaboration}{Collaboration}{2015}]%
        {openscience}
\bibfield{author}{\bibinfo{person}{Open~Science Collaboration}.}
  \bibinfo{year}{2015}\natexlab{}.
\newblock \showarticletitle{Estimating the reproducibility of psychological
  science}.
\newblock \bibinfo{journal}{\emph{Science}} \bibinfo{volume}{349},
  \bibinfo{number}{6251} (\bibinfo{year}{2015}), \bibinfo{pages}{aac4716}.
\newblock
\urldef\tempurl%
\url{https://doi.org/10.1126/science.aac4716}
\showDOI{\tempurl}
\showeprint{https://www.science.org/doi/pdf/10.1126/science.aac4716}


\bibitem[\protect\citeauthoryear{Cordero, Groechel, and Matarić}{Cordero
  et~al\mbox{.}}{2022}]%
        {cordero2022reporting}
\bibfield{author}{\bibinfo{person}{Julia~R Cordero}, \bibinfo{person}{Thomas~R
  Groechel}, {and} \bibinfo{person}{Maja~J Matarić}.}
  \bibinfo{year}{2022}\natexlab{}.
\newblock \showarticletitle{A Review and Recommendations on Reporting
  Recruitment and Compensation Information in HRI Research Papers}. In
  \bibinfo{booktitle}{\emph{2022 31st IEEE International Conference on Robot
  and Human Interactive Communication (RO-MAN)}}. \bibinfo{pages}{1627--1633}.
\newblock
\urldef\tempurl%
\url{https://doi.org/10.1109/RO-MAN53752.2022.9900744}
\showDOI{\tempurl}


\bibitem[\protect\citeauthoryear{Couto, Chandra, Yadollahi, and Charisi}{Couto
  et~al\mbox{.}}{2022}]%
        {couto2022child}
\bibfield{author}{\bibinfo{person}{Marta Couto}, \bibinfo{person}{Shruti
  Chandra}, \bibinfo{person}{Elmira Yadollahi}, {and} \bibinfo{person}{Vicky
  Charisi}.} \bibinfo{year}{2022}\natexlab{}.
\newblock \bibinfo{title}{Child-robot interaction: Design, evaluation, and
  novel solutions}.
\newblock , \bibinfo{numpages}{151--156}~pages.
\newblock


\bibitem[\protect\citeauthoryear{de~Souza~Jeronimo, de~Albuquerque~Wheler,
  de~Oliveira, Melo, Bastos-Filho, and Kelner}{de~Souza~Jeronimo
  et~al\mbox{.}}{2022}]%
        {de2022comparing}
\bibfield{author}{\bibinfo{person}{Bruno de Souza~Jeronimo},
  \bibinfo{person}{Anna~Priscilla de Albuquerque~Wheler},
  \bibinfo{person}{Jos{\'e} Paulo~G de Oliveira}, \bibinfo{person}{Rodrigo
  Melo}, \bibinfo{person}{Carmelo~JA Bastos-Filho}, {and}
  \bibinfo{person}{Judith Kelner}.} \bibinfo{year}{2022}\natexlab{}.
\newblock \showarticletitle{Comparing Social Robot Embodiment for Child Musical
  Education}.
\newblock \bibinfo{journal}{\emph{Journal of Intelligent \& Robotic Systems}}
  \bibinfo{volume}{105}, \bibinfo{number}{2} (\bibinfo{year}{2022}),
  \bibinfo{pages}{28}.
\newblock


\bibitem[\protect\citeauthoryear{DiPaola, Charisi, Breazeal, and
  Sabanovic}{DiPaola et~al\mbox{.}}{2023}]%
        {dipaola2023fundamental}
\bibfield{author}{\bibinfo{person}{Daniella DiPaola}, \bibinfo{person}{Vicky
  Charisi}, \bibinfo{person}{Cynthia Breazeal}, {and} \bibinfo{person}{Selma
  Sabanovic}.} \bibinfo{year}{2023}\natexlab{}.
\newblock \showarticletitle{Children's Fundamental Rights in Human-Robot
  Interaction Research: A Systematic Review}. In
  \bibinfo{booktitle}{\emph{Companion of the 2023 ACM/IEEE International
  Conference on Human-Robot Interaction}} (Stockholm, Sweden)
  \emph{(\bibinfo{series}{HRI '23})}. \bibinfo{publisher}{Association for
  Computing Machinery}, \bibinfo{address}{New York, NY, USA},
  \bibinfo{pages}{561–566}.
\newblock
\showISBNx{9781450399708}
\urldef\tempurl%
\url{https://doi.org/10.1145/3568294.3580148}
\showDOI{\tempurl}


\bibitem[\protect\citeauthoryear{Fraune, Leite, Karatas, Amirova, Legeleux,
  Sandygulova, Neerincx, Dilip~Tikas, Gunes, Mohan, Abbasi, Shenoy,
  Scassellati, de~Visser, and Komatsu}{Fraune et~al\mbox{.}}{2022}]%
        {fraune2022lessons}
\bibfield{author}{\bibinfo{person}{Marlena~R. Fraune}, \bibinfo{person}{Iolanda
  Leite}, \bibinfo{person}{Nihan Karatas}, \bibinfo{person}{Aida Amirova},
  \bibinfo{person}{Amélie Legeleux}, \bibinfo{person}{Anara Sandygulova},
  \bibinfo{person}{Anouk Neerincx}, \bibinfo{person}{Gaurav Dilip~Tikas},
  \bibinfo{person}{Hatice Gunes}, \bibinfo{person}{Mayumi Mohan},
  \bibinfo{person}{Nida~Itrat Abbasi}, \bibinfo{person}{Sudhir Shenoy},
  \bibinfo{person}{Brian Scassellati}, \bibinfo{person}{Ewart~J. de Visser},
  {and} \bibinfo{person}{Takanori Komatsu}.} \bibinfo{year}{2022}\natexlab{}.
\newblock \showarticletitle{Lessons Learned About Designing and Conducting
  Studies From HRI Experts}.
\newblock \bibinfo{journal}{\emph{Frontiers in Robotics and AI}}
  \bibinfo{volume}{8} (\bibinfo{year}{2022}).
\newblock
\showISSN{2296-9144}
\urldef\tempurl%
\url{https://doi.org/10.3389/frobt.2021.772141}
\showDOI{\tempurl}


\bibitem[\protect\citeauthoryear{Gunes, Broz, Crawford, der P{\"u}tten, Strait,
  and Riek}{Gunes et~al\mbox{.}}{2022}]%
        {gunes2022reproducibility}
\bibfield{author}{\bibinfo{person}{Hatice Gunes}, \bibinfo{person}{Frank Broz},
  \bibinfo{person}{Chris~S Crawford}, \bibinfo{person}{Astrid Rosenthal-von der
  P{\"u}tten}, \bibinfo{person}{Megan Strait}, {and} \bibinfo{person}{Laurel
  Riek}.} \bibinfo{year}{2022}\natexlab{}.
\newblock \showarticletitle{Reproducibility in Human-Robot Interaction:
  Furthering the Science of HRI}.
\newblock \bibinfo{journal}{\emph{Current Robotics Reports}}
  (\bibinfo{year}{2022}), \bibinfo{pages}{1--12}.
\newblock


\bibitem[\protect\citeauthoryear{Hoffman and Zhao}{Hoffman and Zhao}{2020}]%
        {hoffman2020}
\bibfield{author}{\bibinfo{person}{Guy Hoffman} {and} \bibinfo{person}{Xuan
  Zhao}.} \bibinfo{year}{2020}\natexlab{}.
\newblock \showarticletitle{A Primer for Conducting Experiments in
  Human–Robot Interaction}.
\newblock \bibinfo{journal}{\emph{J. Hum.-Robot Interact.}}
  \bibinfo{volume}{10}, \bibinfo{number}{1}, Article \bibinfo{articleno}{6}
  (\bibinfo{date}{oct} \bibinfo{year}{2020}), \bibinfo{numpages}{31}~pages.
\newblock
\urldef\tempurl%
\url{https://doi.org/10.1145/3412374}
\showDOI{\tempurl}


\bibitem[\protect\citeauthoryear{Irfan, Kennedy, Lemaignan, Papadopoulos,
  Senft, and Belpaeme}{Irfan et~al\mbox{.}}{2018}]%
        {irfan2018social}
\bibfield{author}{\bibinfo{person}{Bahar Irfan}, \bibinfo{person}{James
  Kennedy}, \bibinfo{person}{S{\'e}verin Lemaignan}, \bibinfo{person}{Fotios
  Papadopoulos}, \bibinfo{person}{Emmanuel Senft}, {and} \bibinfo{person}{Tony
  Belpaeme}.} \bibinfo{year}{2018}\natexlab{}.
\newblock \showarticletitle{Social psychology and human-robot interaction: An
  uneasy marriage}. In \bibinfo{booktitle}{\emph{Companion of the 2018 ACM/IEEE
  International Conference on Human-Robot Interaction}}.
  \bibinfo{pages}{13--20}.
\newblock


\bibitem[\protect\citeauthoryear{Ismail, Hanapiah, Belpaeme, Dambre, and
  Wyffels}{Ismail et~al\mbox{.}}{2021}]%
        {ismail2021analysis}
\bibfield{author}{\bibinfo{person}{Luthffi~Idzhar Ismail},
  \bibinfo{person}{Fazah~Akhtar Hanapiah}, \bibinfo{person}{Tony Belpaeme},
  \bibinfo{person}{Joni Dambre}, {and} \bibinfo{person}{Francis Wyffels}.}
  \bibinfo{year}{2021}\natexlab{}.
\newblock \showarticletitle{Analysis of attention in child--robot interaction
  among children diagnosed with cognitive impairment}.
\newblock \bibinfo{journal}{\emph{International Journal of Social Robotics}}
  \bibinfo{volume}{13} (\bibinfo{year}{2021}), \bibinfo{pages}{141--152}.
\newblock


\bibitem[\protect\citeauthoryear{Jost, Pévédic, Belpaeme, Bethel,
  Chrysostomou, Crook, Grandgeorge, and Mirnig}{Jost et~al\mbox{.}}{2020}]%
        {2020jost}
\bibfield{editor}{\bibinfo{person}{Céline Jost}, \bibinfo{person}{Brigitte~Le
  Pévédic}, \bibinfo{person}{Tony Belpaeme}, \bibinfo{person}{Cindy Bethel},
  \bibinfo{person}{Dimitrios Chrysostomou}, \bibinfo{person}{Nigel Crook},
  \bibinfo{person}{Marine Grandgeorge}, {and} \bibinfo{person}{Nicole Mirnig}}
  (Eds.). \bibinfo{year}{2020}\natexlab{}.
\newblock \bibinfo{booktitle}{\emph{Human-Robot Interaction}}.
  Vol.~\bibinfo{volume}{12}.
\newblock \bibinfo{publisher}{Springer International Publishing}.
\newblock
\showISBNx{978-3-030-42306-3}
\urldef\tempurl%
\url{https://doi.org/10.1007/978-3-030-42307-0}
\showDOI{\tempurl}


\bibitem[\protect\citeauthoryear{Kewalramani and Veresov}{Kewalramani and
  Veresov}{2022}]%
        {kewalramani2022multimodal}
\bibfield{author}{\bibinfo{person}{Sarika Kewalramani} {and}
  \bibinfo{person}{Nikolay Veresov}.} \bibinfo{year}{2022}\natexlab{}.
\newblock \showarticletitle{Multimodal creative inquiry: Theorising a new
  approach for children’s Science meaning-making in Early Childhood
  Education}.
\newblock \bibinfo{journal}{\emph{Research in Science Education}}
  (\bibinfo{year}{2022}), \bibinfo{pages}{1--21}.
\newblock


\bibitem[\protect\citeauthoryear{Konijn, Jansen, Mondaca~Bustos, Hobbelink, and
  Preciado~Vanegas}{Konijn et~al\mbox{.}}{2022}]%
        {konijn2022social}
\bibfield{author}{\bibinfo{person}{Elly~A Konijn}, \bibinfo{person}{Brechtje
  Jansen}, \bibinfo{person}{Victoria Mondaca~Bustos},
  \bibinfo{person}{Veerle~LNF Hobbelink}, {and} \bibinfo{person}{Daniel
  Preciado~Vanegas}.} \bibinfo{year}{2022}\natexlab{}.
\newblock \showarticletitle{Social robots for (second) language learning in
  (migrant) primary school children}.
\newblock \bibinfo{journal}{\emph{International Journal of Social Robotics}}
  (\bibinfo{year}{2022}), \bibinfo{pages}{1--17}.
\newblock


\bibitem[\protect\citeauthoryear{Koole and Lakens}{Koole and Lakens}{2012}]%
        {koole2012replications}
\bibfield{author}{\bibinfo{person}{Sander~L. Koole} {and}
  \bibinfo{person}{Daniël Lakens}.} \bibinfo{year}{2012}\natexlab{}.
\newblock \showarticletitle{Rewarding Replications: A Sure and Simple Way to
  Improve Psychological Science}.
\newblock \bibinfo{journal}{\emph{Perspectives on Psychological Science}}
  \bibinfo{volume}{7}, \bibinfo{number}{6} (\bibinfo{year}{2012}),
  \bibinfo{pages}{608--614}.
\newblock
\urldef\tempurl%
\url{https://doi.org/10.1177/1745691612462586}
\showDOI{\tempurl}
\showeprint{https://doi.org/10.1177/1745691612462586}
\newblock
\shownote{PMID: 26168120.}


\bibitem[\protect\citeauthoryear{Langer, Marshall, and Levy-Tzedek}{Langer
  et~al\mbox{.}}{2023}]%
        {langer2023ethical}
\bibfield{author}{\bibinfo{person}{Allison Langer}, \bibinfo{person}{Peter~J
  Marshall}, {and} \bibinfo{person}{Shelly Levy-Tzedek}.}
  \bibinfo{year}{2023}\natexlab{}.
\newblock \showarticletitle{Ethical Considerations in Child-Robot
  Interactions}.
\newblock \bibinfo{journal}{\emph{Neuroscience \& Biobehavioral Reviews}}
  (\bibinfo{year}{2023}), \bibinfo{pages}{105230}.
\newblock


\bibitem[\protect\citeauthoryear{Leichtmann, Nitsch, and Mara}{Leichtmann
  et~al\mbox{.}}{2022}]%
        {leichtmann2022crisis}
\bibfield{author}{\bibinfo{person}{Benedikt Leichtmann},
  \bibinfo{person}{Verena Nitsch}, {and} \bibinfo{person}{Martina Mara}.}
  \bibinfo{year}{2022}\natexlab{}.
\newblock \showarticletitle{Crisis ahead? Why human-robot interaction user
  studies may have replicability problems and directions for improvement}.
\newblock \bibinfo{journal}{\emph{Frontiers in Robotics and AI}}
  \bibinfo{volume}{9} (\bibinfo{year}{2022}).
\newblock


\bibitem[\protect\citeauthoryear{Liu, Guo, et~al\mbox{.}}{Liu
  et~al\mbox{.}}{2022}]%
        {liu2022auxiliary}
\bibfield{author}{\bibinfo{person}{Kai Liu}, \bibinfo{person}{Cheng Guo},
  {et~al\mbox{.}}} \bibinfo{year}{2022}\natexlab{}.
\newblock \showarticletitle{The Auxiliary Function of Intelligent Mobile Robot
  in the Standard Training of Children’s Sports Movement}.
\newblock \bibinfo{journal}{\emph{Wireless Communications and Mobile
  Computing}}  \bibinfo{volume}{2022} (\bibinfo{year}{2022}).
\newblock


\bibitem[\protect\citeauthoryear{Livingstone, Mascheroni, and
  Staksrud}{Livingstone et~al\mbox{.}}{2018}]%
        {livingstone2018european}
\bibfield{author}{\bibinfo{person}{Sonia Livingstone},
  \bibinfo{person}{Giovanna Mascheroni}, {and} \bibinfo{person}{Elisabeth
  Staksrud}.} \bibinfo{year}{2018}\natexlab{}.
\newblock \showarticletitle{European research on children’s internet use:
  Assessing the past and anticipating the future}.
\newblock \bibinfo{journal}{\emph{New media \& society}} \bibinfo{volume}{20},
  \bibinfo{number}{3} (\bibinfo{year}{2018}), \bibinfo{pages}{1103--1122}.
\newblock


\bibitem[\protect\citeauthoryear{Lopez-Caudana, Ramirez-Montoya,
  Mart{\'\i}nez-P{\'e}rez, and Rodr{\'\i}guez-Abitia}{Lopez-Caudana
  et~al\mbox{.}}{2020}]%
        {lopez2020using}
\bibfield{author}{\bibinfo{person}{Edgar Lopez-Caudana},
  \bibinfo{person}{Maria~Soledad Ramirez-Montoya}, \bibinfo{person}{Sandra
  Mart{\'\i}nez-P{\'e}rez}, {and} \bibinfo{person}{Guillermo
  Rodr{\'\i}guez-Abitia}.} \bibinfo{year}{2020}\natexlab{}.
\newblock \showarticletitle{Using robotics to enhance active learning in
  mathematics: A multi-scenario study}.
\newblock \bibinfo{journal}{\emph{Mathematics}} \bibinfo{volume}{8},
  \bibinfo{number}{12} (\bibinfo{year}{2020}), \bibinfo{pages}{2163}.
\newblock


\bibitem[\protect\citeauthoryear{L{\"u}cking, Lier, Bernotat, Wachsmuth,
  {\^S}abanovi{\'c}, and Eyssel}{L{\"u}cking et~al\mbox{.}}{2018}]%
        {lucking2018geographically}
\bibfield{author}{\bibinfo{person}{Phillip L{\"u}cking},
  \bibinfo{person}{Florian Lier}, \bibinfo{person}{Jasmin Bernotat},
  \bibinfo{person}{Sven Wachsmuth}, \bibinfo{person}{Selma {\^S}abanovi{\'c}},
  {and} \bibinfo{person}{Friederike Eyssel}.} \bibinfo{year}{2018}\natexlab{}.
\newblock \showarticletitle{Geographically distributed deployment of
  reproducible hri experiments in an interdisciplinary research context}. In
  \bibinfo{booktitle}{\emph{Companion of the 2018 ACM/IEEE International
  Conference on Human-Robot Interaction}}. \bibinfo{pages}{181--182}.
\newblock


\bibitem[\protect\citeauthoryear{Martin, MacIntyre, Perry, Clift, Pedell, and
  Kaufman}{Martin et~al\mbox{.}}{2020}]%
        {martin2020young}
\bibfield{author}{\bibinfo{person}{Dorothea~U Martin},
  \bibinfo{person}{Madeline~I MacIntyre}, \bibinfo{person}{Conrad Perry},
  \bibinfo{person}{Georgia Clift}, \bibinfo{person}{Sonja Pedell}, {and}
  \bibinfo{person}{Jordy Kaufman}.} \bibinfo{year}{2020}\natexlab{}.
\newblock \showarticletitle{Young children’s indiscriminate helping behavior
  toward a humanoid robot}.
\newblock \bibinfo{journal}{\emph{Frontiers in psychology}}
  \bibinfo{volume}{11} (\bibinfo{year}{2020}), \bibinfo{pages}{239}.
\newblock


\bibitem[\protect\citeauthoryear{Moher, Liberati, Tetzlaff, Altman, and
  PRISMA~Group*}{Moher et~al\mbox{.}}{2009}]%
        {moher2009preferred}
\bibfield{author}{\bibinfo{person}{David Moher}, \bibinfo{person}{Alessandro
  Liberati}, \bibinfo{person}{Jennifer Tetzlaff}, \bibinfo{person}{Douglas~G
  Altman}, {and} \bibinfo{person}{the PRISMA~Group*}.}
  \bibinfo{year}{2009}\natexlab{}.
\newblock \showarticletitle{Preferred reporting items for systematic reviews
  and meta-analyses: the PRISMA statement}.
\newblock \bibinfo{journal}{\emph{Annals of internal medicine}}
  \bibinfo{volume}{151}, \bibinfo{number}{4} (\bibinfo{year}{2009}),
  \bibinfo{pages}{264--269}.
\newblock


\bibitem[\protect\citeauthoryear{Mott, Bejarano, and Williams}{Mott
  et~al\mbox{.}}{2022}]%
        {mott2022robot}
\bibfield{author}{\bibinfo{person}{Terran Mott}, \bibinfo{person}{Alexandra
  Bejarano}, {and} \bibinfo{person}{Tom Williams}.}
  \bibinfo{year}{2022}\natexlab{}.
\newblock \showarticletitle{Robot co-design can help us engage child
  stakeholders in ethical reflection}. In \bibinfo{booktitle}{\emph{2022 17th
  ACM/IEEE International Conference on Human-Robot Interaction (HRI)}}. IEEE,
  \bibinfo{pages}{14--23}.
\newblock


\bibitem[\protect\citeauthoryear{Munaf{\`o}, Nosek, Bishop, Button, Chambers,
  Percie~du Sert, Simonsohn, Wagenmakers, Ware, and Ioannidis}{Munaf{\`o}
  et~al\mbox{.}}{2017}]%
        {munafo2017manifesto}
\bibfield{author}{\bibinfo{person}{Marcus~R Munaf{\`o}},
  \bibinfo{person}{Brian~A Nosek}, \bibinfo{person}{Dorothy~VM Bishop},
  \bibinfo{person}{Katherine~S Button}, \bibinfo{person}{Christopher~D
  Chambers}, \bibinfo{person}{Nathalie Percie~du Sert}, \bibinfo{person}{Uri
  Simonsohn}, \bibinfo{person}{Eric-Jan Wagenmakers},
  \bibinfo{person}{Jennifer~J Ware}, {and} \bibinfo{person}{John Ioannidis}.}
  \bibinfo{year}{2017}\natexlab{}.
\newblock \showarticletitle{A manifesto for reproducible science}.
\newblock \bibinfo{journal}{\emph{Nature human behaviour}} \bibinfo{volume}{1},
  \bibinfo{number}{1} (\bibinfo{year}{2017}), \bibinfo{pages}{1--9}.
\newblock


\bibitem[\protect\citeauthoryear{Neerincx, Hiwat, and De~Graaf}{Neerincx
  et~al\mbox{.}}{2021}]%
        {neerincx2021social}
\bibfield{author}{\bibinfo{person}{Anouk Neerincx}, \bibinfo{person}{Thirza
  Hiwat}, {and} \bibinfo{person}{Maartje De~Graaf}.}
  \bibinfo{year}{2021}\natexlab{}.
\newblock \showarticletitle{Social robot for health check and entertainment in
  waiting room: Child’s engagement and parent’s involvement}. In
  \bibinfo{booktitle}{\emph{Adjunct Proceedings of the 29th ACM Conference on
  User Modeling, Adaptation and Personalization}}. \bibinfo{pages}{120--125}.
\newblock


\bibitem[\protect\citeauthoryear{Nightingale}{Nightingale}{2009}]%
        {nightingale2009guide}
\bibfield{author}{\bibinfo{person}{Alison Nightingale}.}
  \bibinfo{year}{2009}\natexlab{}.
\newblock \showarticletitle{A guide to systematic literature reviews}.
\newblock \bibinfo{journal}{\emph{Surgery (Oxford)}} \bibinfo{volume}{27},
  \bibinfo{number}{9} (\bibinfo{year}{2009}), \bibinfo{pages}{381--384}.
\newblock


\bibitem[\protect\citeauthoryear{Nijssen, M{\"u}ller, Bosse, and
  Paulus}{Nijssen et~al\mbox{.}}{2021}]%
        {nijssen2021you}
\bibfield{author}{\bibinfo{person}{Sari~RR Nijssen},
  \bibinfo{person}{Barbara~CN M{\"u}ller}, \bibinfo{person}{Tibor Bosse}, {and}
  \bibinfo{person}{Markus Paulus}.} \bibinfo{year}{2021}\natexlab{}.
\newblock \showarticletitle{You, robot? The role of anthropomorphic emotion
  attributions in children’s sharing with a robot}.
\newblock \bibinfo{journal}{\emph{International Journal of Child-Computer
  Interaction}}  \bibinfo{volume}{30} (\bibinfo{year}{2021}),
  \bibinfo{pages}{100319}.
\newblock


\bibitem[\protect\citeauthoryear{Nosek, Alter, Banks, Borsboom, Bowman,
  Breckler, Buck, Chambers, Chin, Christensen, et~al\mbox{.}}{Nosek
  et~al\mbox{.}}{2015}]%
        {nosek2015promoting}
\bibfield{author}{\bibinfo{person}{Brian~A Nosek}, \bibinfo{person}{George
  Alter}, \bibinfo{person}{George~C Banks}, \bibinfo{person}{Denny Borsboom},
  \bibinfo{person}{Sara~D Bowman}, \bibinfo{person}{Steven~J Breckler},
  \bibinfo{person}{Stuart Buck}, \bibinfo{person}{Christopher~D Chambers},
  \bibinfo{person}{Gilbert Chin}, \bibinfo{person}{Garret Christensen},
  {et~al\mbox{.}}} \bibinfo{year}{2015}\natexlab{}.
\newblock \showarticletitle{Promoting an open research culture}.
\newblock \bibinfo{journal}{\emph{Science}} \bibinfo{volume}{348},
  \bibinfo{number}{6242} (\bibinfo{year}{2015}), \bibinfo{pages}{1422--1425}.
\newblock


\bibitem[\protect\citeauthoryear{Nosek, Hardwicke, Moshontz, Allard, Corker,
  Dreber, Fidler, Hilgard, Kline~Struhl, Nuijten, et~al\mbox{.}}{Nosek
  et~al\mbox{.}}{2022}]%
        {nosek2022replicability}
\bibfield{author}{\bibinfo{person}{Brian~A Nosek}, \bibinfo{person}{Tom~E
  Hardwicke}, \bibinfo{person}{Hannah Moshontz}, \bibinfo{person}{Aur{\'e}lien
  Allard}, \bibinfo{person}{Katherine~S Corker}, \bibinfo{person}{Anna Dreber},
  \bibinfo{person}{Fiona Fidler}, \bibinfo{person}{Joe Hilgard},
  \bibinfo{person}{Melissa Kline~Struhl}, \bibinfo{person}{Mich{\`e}le~B
  Nuijten}, {et~al\mbox{.}}} \bibinfo{year}{2022}\natexlab{}.
\newblock \showarticletitle{Replicability, robustness, and reproducibility in
  psychological science}.
\newblock \bibinfo{journal}{\emph{Annual review of psychology}}
  \bibinfo{volume}{73} (\bibinfo{year}{2022}), \bibinfo{pages}{719--748}.
\newblock


\bibitem[\protect\citeauthoryear{Oranc and K{\"u}ntay}{Oranc and
  K{\"u}ntay}{2020}]%
        {oranc2020children}
\bibfield{author}{\bibinfo{person}{Cansu Oranc} {and} \bibinfo{person}{Aylin~C
  K{\"u}ntay}.} \bibinfo{year}{2020}\natexlab{}.
\newblock \showarticletitle{Children’s perception of social robots as a
  source of information across different domains of knowledge}.
\newblock \bibinfo{journal}{\emph{Cognitive Development}}  \bibinfo{volume}{54}
  (\bibinfo{year}{2020}), \bibinfo{pages}{100875}.
\newblock


\bibitem[\protect\citeauthoryear{Riek}{Riek}{2012}]%
        {riek2022wizard}
\bibfield{author}{\bibinfo{person}{Laurel~D. Riek}.}
  \bibinfo{year}{2012}\natexlab{}.
\newblock \showarticletitle{Wizard of Oz Studies in HRI: A Systematic Review
  and New Reporting Guidelines}.
\newblock \bibinfo{journal}{\emph{J. Hum.-Robot Interact.}}
  \bibinfo{volume}{1}, \bibinfo{number}{1} (\bibinfo{date}{jul}
  \bibinfo{year}{2012}), \bibinfo{pages}{119–136}.
\newblock
\urldef\tempurl%
\url{https://doi.org/10.5898/JHRI.1.1.Riek}
\showDOI{\tempurl}


\bibitem[\protect\citeauthoryear{Romero}{Romero}{2019}]%
        {romero2019phylo}
\bibfield{author}{\bibinfo{person}{F Romero}.} \bibinfo{year}{2019}\natexlab{}.
\newblock \showarticletitle{Philosophy of Science and the Replicability
  Crisis}. In \bibinfo{booktitle}{\emph{Philos. Compass}}.
\newblock
\urldef\tempurl%
\url{https://doi.org/10.1111/phc3.12633}
\showDOI{\tempurl}


\bibitem[\protect\citeauthoryear{Scheel, Tiokhin, Isager, and Lakens}{Scheel
  et~al\mbox{.}}{2021}]%
        {scheel2021hypothesis}
\bibfield{author}{\bibinfo{person}{Anne~M. Scheel}, \bibinfo{person}{Leonid
  Tiokhin}, \bibinfo{person}{Peder~M. Isager}, {and} \bibinfo{person}{Daniël
  Lakens}.} \bibinfo{year}{2021}\natexlab{}.
\newblock \showarticletitle{Why Hypothesis Testers Should Spend Less Time
  Testing Hypotheses}.
\newblock \bibinfo{journal}{\emph{Perspectives on Psychological Science}}
  \bibinfo{volume}{16}, \bibinfo{number}{4} (\bibinfo{year}{2021}),
  \bibinfo{pages}{744--755}.
\newblock
\urldef\tempurl%
\url{https://doi.org/10.1177/1745691620966795}
\showDOI{\tempurl}
\showeprint{https://doi.org/10.1177/1745691620966795}
\newblock
\shownote{PMID: 33326363.}


\bibitem[\protect\citeauthoryear{Serholt, Pareto, Ekstr{\"o}m, and
  Ljungblad}{Serholt et~al\mbox{.}}{2020}]%
        {serholt2020trouble}
\bibfield{author}{\bibinfo{person}{Sofia Serholt}, \bibinfo{person}{Lena
  Pareto}, \bibinfo{person}{Sara Ekstr{\"o}m}, {and} \bibinfo{person}{Sara
  Ljungblad}.} \bibinfo{year}{2020}\natexlab{}.
\newblock \showarticletitle{Trouble and Repair in Child--Robot Interaction: a
  study of complex interactions with a robot tutee in a primary school
  classroom}.
\newblock \bibinfo{journal}{\emph{Frontiers in Robotics and AI}}
  \bibinfo{volume}{7} (\bibinfo{year}{2020}), \bibinfo{pages}{46}.
\newblock


\bibitem[\protect\citeauthoryear{Shahab, Taheri, Mokhtari, Shariati, Heidari,
  Meghdari, and Alemi}{Shahab et~al\mbox{.}}{2022}]%
        {shahab2022utilizing}
\bibfield{author}{\bibinfo{person}{Mojtaba Shahab}, \bibinfo{person}{Alireza
  Taheri}, \bibinfo{person}{Mohammad Mokhtari}, \bibinfo{person}{Azadeh
  Shariati}, \bibinfo{person}{Rozita Heidari}, \bibinfo{person}{Ali Meghdari},
  {and} \bibinfo{person}{Minoo Alemi}.} \bibinfo{year}{2022}\natexlab{}.
\newblock \showarticletitle{Utilizing social virtual reality robot (V2R) for
  music education to children with high-functioning autism}.
\newblock \bibinfo{journal}{\emph{Education and Information Technologies}}
  (\bibinfo{year}{2022}), \bibinfo{pages}{1--25}.
\newblock


\bibitem[\protect\citeauthoryear{Spitale, Silleresi, Cosentino, Panzeri, and
  Garzotto}{Spitale et~al\mbox{.}}{2020}]%
        {spitale2020whom}
\bibfield{author}{\bibinfo{person}{Micol Spitale}, \bibinfo{person}{Silvia
  Silleresi}, \bibinfo{person}{Giulia Cosentino}, \bibinfo{person}{Francesca
  Panzeri}, {and} \bibinfo{person}{Franca Garzotto}.}
  \bibinfo{year}{2020}\natexlab{}.
\newblock \showarticletitle{" Whom would you like to talk with?" exploring
  conversational agents for children's linguistic assessment}. In
  \bibinfo{booktitle}{\emph{Proceedings of the interaction design and children
  conference}}. \bibinfo{pages}{262--272}.
\newblock


\bibitem[\protect\citeauthoryear{Stower, Ligthart, Van~Straten, Calvo-Barajas,
  Velner, and Beelen}{Stower et~al\mbox{.}}{2021}]%
        {stower2021interdisciplinary}
\bibfield{author}{\bibinfo{person}{Rebecca Stower}, \bibinfo{person}{Mike
  Ligthart}, \bibinfo{person}{Caroline Van~Straten}, \bibinfo{person}{Natalia
  Calvo-Barajas}, \bibinfo{person}{Ella Velner}, {and} \bibinfo{person}{Thomas
  Beelen}.} \bibinfo{year}{2021}\natexlab{}.
\newblock \showarticletitle{Interdisciplinary research methods for child-robot
  relationship formation}. In \bibinfo{booktitle}{\emph{Companion of the 2021
  ACM/IEEE International Conference on Human-Robot Interaction}}.
  \bibinfo{pages}{700--702}.
\newblock


\bibitem[\protect\citeauthoryear{Stower, Ligthart, Spitale, Calvo-Barajas, and
  de~Droog}{Stower et~al\mbox{.}}{2023}]%
        {stower2023workshop}
\bibfield{author}{\bibinfo{person}{Rebecca Stower}, \bibinfo{person}{Mike~E.U.
  Ligthart}, \bibinfo{person}{Micol Spitale}, \bibinfo{person}{Natalia
  Calvo-Barajas}, {and} \bibinfo{person}{Simone~M. de Droog}.}
  \bibinfo{year}{2023}\natexlab{}.
\newblock \showarticletitle{CRITTER: Child-Robot Interaction and
  Interdisciplinary Research}. In \bibinfo{booktitle}{\emph{Companion of the
  2023 ACM/IEEE International Conference on Human-Robot Interaction}}
  (Stockholm, Sweden) \emph{(\bibinfo{series}{HRI '23})}.
  \bibinfo{publisher}{Association for Computing Machinery},
  \bibinfo{address}{New York, NY, USA}, \bibinfo{pages}{926–928}.
\newblock
\showISBNx{9781450399708}
\urldef\tempurl%
\url{https://doi.org/10.1145/3568294.3579955}
\showDOI{\tempurl}


\bibitem[\protect\citeauthoryear{Strait, Lier, Bernotat, Wachsmuth, Eyssel,
  Goldstone, and {\v{S}}abanovi{\'c}}{Strait et~al\mbox{.}}{2020}]%
        {strait2020three}
\bibfield{author}{\bibinfo{person}{Megan Strait}, \bibinfo{person}{Florian
  Lier}, \bibinfo{person}{Jasmin Bernotat}, \bibinfo{person}{Sven Wachsmuth},
  \bibinfo{person}{Friederike Eyssel}, \bibinfo{person}{Robert Goldstone},
  {and} \bibinfo{person}{Selma {\v{S}}abanovi{\'c}}.}
  \bibinfo{year}{2020}\natexlab{}.
\newblock \showarticletitle{A three-site reproduction of the joint simon effect
  with the nao robot}. In \bibinfo{booktitle}{\emph{Proceedings of the 2020
  ACM/IEEE International Conference on Human-Robot Interaction}}.
  \bibinfo{pages}{103--111}.
\newblock


\bibitem[\protect\citeauthoryear{Tian, Lubold, Friedman, and Walker}{Tian
  et~al\mbox{.}}{2020}]%
        {tian2020understanding}
\bibfield{author}{\bibinfo{person}{Xiaoyi Tian}, \bibinfo{person}{Nichola
  Lubold}, \bibinfo{person}{Leah Friedman}, {and} \bibinfo{person}{Erin
  Walker}.} \bibinfo{year}{2020}\natexlab{}.
\newblock \showarticletitle{Understanding Rapport over Multiple Sessions with a
  Social, Teachable Robot}. In \bibinfo{booktitle}{\emph{Artificial
  Intelligence in Education: 21st International Conference, AIED 2020, Ifrane,
  Morocco, July 6--10, 2020, Proceedings, Part II 21}}. Springer,
  \bibinfo{pages}{318--323}.
\newblock


\bibitem[\protect\citeauthoryear{Tolksdorf, Crawshaw, and Rohlfing}{Tolksdorf
  et~al\mbox{.}}{2021a}]%
        {tolksdorf2021comparing}
\bibfield{author}{\bibinfo{person}{Nils~F Tolksdorf},
  \bibinfo{person}{Camilla~E Crawshaw}, {and} \bibinfo{person}{Katharina~J
  Rohlfing}.} \bibinfo{year}{2021}\natexlab{a}.
\newblock \showarticletitle{Comparing the effects of a different social partner
  (social robot vs. human) on children's social referencing in interaction}. In
  \bibinfo{booktitle}{\emph{Frontiers in Education}}, Vol.~\bibinfo{volume}{5}.
  Frontiers Media SA, \bibinfo{pages}{569615}.
\newblock


\bibitem[\protect\citeauthoryear{Tolksdorf, Siebert, Zorn, Horwath, and
  Rohlfing}{Tolksdorf et~al\mbox{.}}{2021b}]%
        {tolksdorf2021ethical}
\bibfield{author}{\bibinfo{person}{Nils~F Tolksdorf}, \bibinfo{person}{Scarlet
  Siebert}, \bibinfo{person}{Isabel Zorn}, \bibinfo{person}{Ilona Horwath},
  {and} \bibinfo{person}{Katharina~J Rohlfing}.}
  \bibinfo{year}{2021}\natexlab{b}.
\newblock \showarticletitle{Ethical considerations of applying robots in
  kindergarten settings: Towards an approach from a macroperspective}.
\newblock \bibinfo{journal}{\emph{International Journal of Social Robotics}}
  \bibinfo{volume}{13} (\bibinfo{year}{2021}), \bibinfo{pages}{129--140}.
\newblock


\bibitem[\protect\citeauthoryear{Ullman, Aladia, and Malle}{Ullman
  et~al\mbox{.}}{2021}]%
        {ullman2021challenges}
\bibfield{author}{\bibinfo{person}{Daniel Ullman}, \bibinfo{person}{Salomi
  Aladia}, {and} \bibinfo{person}{Bertram~F Malle}.}
  \bibinfo{year}{2021}\natexlab{}.
\newblock \showarticletitle{Challenges and opportunities for replication
  science in hri: A case study in human-robot trust}. In
  \bibinfo{booktitle}{\emph{Proceedings of the 2021 ACM/IEEE International
  Conference on Human-Robot Interaction}}. \bibinfo{pages}{110--118}.
\newblock


\bibitem[\protect\citeauthoryear{Veling and McGinn}{Veling and McGinn}{2021}]%
        {veling2021qualitative}
\bibfield{author}{\bibinfo{person}{Louise Veling} {and} \bibinfo{person}{Conor
  McGinn}.} \bibinfo{year}{2021}\natexlab{}.
\newblock \showarticletitle{Qualitative research in HRI: A review and
  taxonomy}.
\newblock \bibinfo{journal}{\emph{International Journal of Social Robotics}}
  \bibinfo{volume}{13}, \bibinfo{number}{7} (\bibinfo{year}{2021}),
  \bibinfo{pages}{1689--1709}.
\newblock


\bibitem[\protect\citeauthoryear{Wang, Park, Itakura, Henderson, Kanda,
  Furuhata, and Ishiguro}{Wang et~al\mbox{.}}{2020}]%
        {wang2020infants}
\bibfield{author}{\bibinfo{person}{Ying Wang}, \bibinfo{person}{Yun-Hee Park},
  \bibinfo{person}{Shoji Itakura}, \bibinfo{person}{Annette Margaret~Elizabeth
  Henderson}, \bibinfo{person}{Takayuki Kanda}, \bibinfo{person}{Naoki
  Furuhata}, {and} \bibinfo{person}{Hiroshi Ishiguro}.}
  \bibinfo{year}{2020}\natexlab{}.
\newblock \showarticletitle{Infants' perceptions of cooperation between a human
  and robot}.
\newblock \bibinfo{journal}{\emph{Infant and Child Development}}
  \bibinfo{volume}{29}, \bibinfo{number}{2} (\bibinfo{year}{2020}),
  \bibinfo{pages}{e2161}.
\newblock


\bibitem[\protect\citeauthoryear{Yadollahi, Chandra, Couto, Lim, and
  Sandygulova}{Yadollahi et~al\mbox{.}}{2021}]%
        {yadollahi2021children}
\bibfield{author}{\bibinfo{person}{Elmira Yadollahi}, \bibinfo{person}{Shruti
  Chandra}, \bibinfo{person}{Marta Couto}, \bibinfo{person}{Angelica Lim},
  {and} \bibinfo{person}{Anara Sandygulova}.} \bibinfo{year}{2021}\natexlab{}.
\newblock \showarticletitle{Children, Robots, and Virtual Agents: Present and
  Future Challenges}. In \bibinfo{booktitle}{\emph{Interaction Design and
  Children}}. \bibinfo{pages}{682--686}.
\newblock


\bibitem[\protect\citeauthoryear{Yadollahi, Couto, Dillenbourg, and
  Paiva}{Yadollahi et~al\mbox{.}}{2022}]%
        {yadollahi2022motivating}
\bibfield{author}{\bibinfo{person}{Elmira Yadollahi}, \bibinfo{person}{Marta
  Couto}, \bibinfo{person}{Pierre Dillenbourg}, {and} \bibinfo{person}{Ana
  Paiva}.} \bibinfo{year}{2022}\natexlab{}.
\newblock \showarticletitle{Motivating Children to Practice Perspective-Taking
  Through Playing Games with Cozmo}. In \bibinfo{booktitle}{\emph{2022 31st
  IEEE International Conference on Robot and Human Interactive Communication
  (RO-MAN)}}. IEEE, \bibinfo{pages}{1482--1489}.
\newblock


\bibitem[\protect\citeauthoryear{Yang, Oh, and Wang}{Yang
  et~al\mbox{.}}{2020}]%
        {yang2020hybrid}
\bibfield{author}{\bibinfo{person}{Dapeng Yang}, \bibinfo{person}{Eung-Soo Oh},
  {and} \bibinfo{person}{Yingchun Wang}.} \bibinfo{year}{2020}\natexlab{}.
\newblock \showarticletitle{Hybrid physical education teaching and curriculum
  design based on a voice interactive artificial intelligence educational
  robot}.
\newblock \bibinfo{journal}{\emph{Sustainability}} \bibinfo{volume}{12},
  \bibinfo{number}{19} (\bibinfo{year}{2020}), \bibinfo{pages}{8000}.
\newblock


\bibitem[\protect\citeauthoryear{Zhang, Breazeal, and Park}{Zhang
  et~al\mbox{.}}{2023}]%
        {zhang2023social}
\bibfield{author}{\bibinfo{person}{Xiajie Zhang}, \bibinfo{person}{Cynthia
  Breazeal}, {and} \bibinfo{person}{Hae~Won Park}.}
  \bibinfo{year}{2023}\natexlab{}.
\newblock \showarticletitle{A Social Robot Reading Partner for Explorative
  Guidance}. In \bibinfo{booktitle}{\emph{Proceedings of the 2023 ACM/IEEE
  International Conference on Human-Robot Interaction}}.
  \bibinfo{pages}{341--349}.
\newblock


\bibitem[\protect\citeauthoryear{Zhexenova, Amirova, Abdikarimova,
  Kudaibergenov, Baimakhan, Tleubayev, Asselborn, Johal, Dillenbourg,
  CohenMiller, et~al\mbox{.}}{Zhexenova et~al\mbox{.}}{2020}]%
        {zhexenova2020comparison}
\bibfield{author}{\bibinfo{person}{Zhanel Zhexenova}, \bibinfo{person}{Aida
  Amirova}, \bibinfo{person}{Manshuk Abdikarimova}, \bibinfo{person}{Kuanysh
  Kudaibergenov}, \bibinfo{person}{Nurakhmet Baimakhan}, \bibinfo{person}{Bolat
  Tleubayev}, \bibinfo{person}{Thibault Asselborn}, \bibinfo{person}{Wafa
  Johal}, \bibinfo{person}{Pierre Dillenbourg}, \bibinfo{person}{Anna
  CohenMiller}, {et~al\mbox{.}}} \bibinfo{year}{2020}\natexlab{}.
\newblock \showarticletitle{A comparison of social robot to tablet and teacher
  in a new script learning context}.
\newblock \bibinfo{journal}{\emph{Frontiers in Robotics and AI}}
  \bibinfo{volume}{7} (\bibinfo{year}{2020}), \bibinfo{pages}{99}.
\newblock


\end{thebibliography}

\end{document}